\documentclass[submission, Phys]{SciPost}

\newcommand{\eq}[1]{\begin{equation}#1\end{equation}}
\newcommand{\dd}{\mathrm{d}}
\newcommand{\ee}{\mathrm{e}}

\newcommand{\sh}{\mathrm{sh}}

\newcommand{\Tr}{\mathrm{Tr}}
\newcommand{\ceff}{c_\textrm{\tiny eff}}

\def \be {\begin{equation}} 
\def \eeq {\end{equation}} 

\def \la {\langle} 
\def \ra {\rangle}

\usepackage{amsmath}
\usepackage{amsfonts}
\usepackage{amssymb}
\usepackage{braket}
\usepackage{bbold}
\usepackage{graphics}
\usepackage{graphicx}
\usepackage{subcaption}
\usepackage{leftidx}
\usepackage{psfrag}
\usepackage{color}
\usepackage{colordvi}
\usepackage{bbm}

\usepackage{hyperref}
\hypersetup{colorlinks,bookmarksopen,bookmarksnumbered,
 citecolor=red,
 linkcolor=blue}

\begin{document}

\begin{center}{\Large \textbf{
Entanglement evolution after a global quench across a conformal defect
}}\end{center}

\begin{center}
Luca Capizzi\textsuperscript{1},
Viktor Eisler\textsuperscript{2}
\end{center}

\begin{center}
{\bf 1} SISSA and INFN Sezione di Trieste, via Bonomea 265, I-34136 Trieste, Italy\\
{\bf 2} Institute of Theoretical and Computational Physics, Graz University of Technology,
Petersgasse 16, A-8010 Graz, Austria
\end{center}

\section*{Abstract}
{\bf
We study the evolution of entanglement after a global quench in a one-dimen\-sional quantum system with a localized impurity. For systems described by a conformal field theory, the entanglement entropy between the two regions separated by the defect grows linearly in time. Introducing the notion of boundary twist fields, we show how the slope of this growth can be related to the effective central charge that emerges in the study of ground-state entropy in the presence of the defect. On the other hand, we also consider a particular lattice realization of the quench in a free-fermion chain with a conformal defect. Starting from a gapped initial state, we obtain the slope via a quasiparticle ansatz and observe small discrepancies between the field theory and lattice results, which persist even in the limit of a vanishing gap.

}

\vspace{10pt}
\noindent\rule{\textwidth}{1pt}
\tableofcontents\thispagestyle{fancy}
\noindent\rule{\textwidth}{1pt}
\vspace{10pt}

\section{Introduction}

The characterization of entanglement in quantum many-body systems has become a vast field of research \cite{Amico-08,ccd-09,Laflorencie-15}. A key result is the discovery of an area law for the entanglement entropy in ground states of local Hamiltonians \cite{ecp-10}, and its logarithmic violation in critical one-dimensional systems \cite{cc-09}, whose features are understood via conformal field theory (CFT). In sharp contrast to the low amount of entanglement observed in the ground-state scenario, the entropy typically shows a rapid growth in the context of non-equilibrium dynamics \cite{pksvm-11,ge-15,cem-16}. In particular, it has been shown that if an initial state is suddenly perturbed by the change of a global parameter in the Hamiltonian, a protocol known as global quench \cite{ef-16,cc-05,cc-16}, the entanglement entropy starts to grow linearly in time. This is followed by a saturation to a volume-law behaviour, suggesting the thermalization of the subsystem.

The physical interpretation of the above mechanism relies on the presence of pairs of quasiparticles, generated after the quench protocol and spreading ballistically across the system. On the level of lattice models, this is indeed realized in integrable systems, where the extensive amount of conservation laws \cite{ef-16} gives rise to quasiparticles with infinite lifetime. One can then identify the asymptotic state of the subsystem with a generalized Gibbs ensemble, entirely determined by the occupations of the quasiparticles, fixed solely by the initial state \cite{vr-16, cef-11,cef-12, cef-12a, sc-14,pvw-17, pk-17}. Moreover, one can argue that the very same occupations and corresponding entropy densities characterize also the dynamics, and together with the quasiparticle velocitites they completely fix the slope of the linear entropy growth \cite{ac-16}.

The description above is known under the name of \textit{quasiparticle picture}. While originally formulated for translational invariant systems, in recent years there has been much work devoted to generalize the setting towards inhomogeneous initial states, see e.g. the recent review \cite{abfpr-21}. In contrast, it is much less clear what happens if the translational invariance is broken by the time evolution operator itself, the simplest example being the presence of a localized defect. On one hand, the integrability of the model can be either preserved \cite{fs-96,ksl-00,ma-06} or destroyed \cite{bpmz-09,Santos-03,ddb-22} by the presence of such defects, making their analytical treatment difficult. On the other hand, a class of free-fermion chains with impurities have been investigated \cite{kl-09,ep-12,lsp-18,glc-20,ge-20,ggs-22,Alba-22,ca-22}, showing nontrivial features in their entanglement dynamics and various physical observables, depending on the details of the defect.

In this work we are interested in the characterization of entanglement growth after a global quench, in a system of two critical half-chains that are coupled together via a conformal defect. We first discuss systems described by a CFT with a conformal interface. Introducing the concept of boundary twist fields and exploiting the power of conformal symmetry, we provide a general prediction for the entropy growth between the two halves
\be
S(t) = \frac{\pi \ceff t}{3\beta},
\label{eq:Sgrowth_def}
\eeq
a relation which was already derived in Ref. \cite{wwr-17} with a different approach.
Here $\beta$ is an inverse temperature characterizing the initial state, and $\ceff \in [0,c]$ is the so-called \textit{effective central charge}, a parameter which depends on the transmission properties of the interface. The crucial result is that $\ceff$ is precisely the same parameter that appears in the ground-state entanglement scaling
\be
S = \frac{\ceff}{6}\log \frac{L}{\epsilon},
\label{eq:Sgs_def}
\eeq
in terms of the half-chain length $L$ and cutoff $\epsilon$. In particular, $\ceff=0$ if the halves are decoupled, while $\ceff = c$ if the system is homogeneous, with its explicit transmission dependence known only for the free boson and fermion CFT \cite{ss-08,bb-15}. Similar results were obtained also for the non-relativistic free Fermi gas \cite{cmv-11,cmv-12}.

On the lattice level, the ground-state entropy scaling \eqref{eq:Sgs_def} has first been investigated numerically for the free-fermion and transverse Ising chains \cite{peschel-05,isl-09,eg-10}, with the analytic solution for $\ceff$ found in \cite{ep-10,pe-12}. Furthermore, it has been demonstrated that, for a particular lattice realization of the conformal defect, the very same $\ceff$ appears in the entropy dynamics after a local quench \cite{ep-12}. Here  we shall extend these studies to the global quench from a gapped initial state and show that, while the analogous relation \eqref{eq:Sgrowth_def} is exact at the CFT level, it holds only approximately for the free-fermion chain. We analyze the small discrepancies between the CFT and lattice results for the slope, and conclude that they arise due to the imperfect thermal description of the initial occupation. Our results are obtained by extending the quasiparticle picture to the evolution across the defect via a proper description of the post-quench occupations. We also consider finite-size effects and discuss the special pattern of entanglement revivals \cite{mac-20} found for large times due to reflections.

We structure this manuscript as follows: in section \ref{sec:CFT} we give a CFT description of the global quench, exploiting the relation between R\'enyi entropies and boundary twist fields. We then move to the lattice part in sec. \ref{sec:lattice} and, after introducing the model, we compute the entropy growth within the quasiparticle picture. We draw our conclusions and provide an outlook in sec.~\ref{sec:Conclusions}, leaving some technical material in three appendices.

\section{CFT results\label{sec:CFT}}

In this section we introduce the CFT approach for the global quench in the presence of a defect. We adapt the formalism of Ref. \cite{cc-16} for the dynamics after the global quench, taking into account the presence of the defect. The calculation of the entanglement entropy is based on the replica trick, giving a field theoretic description of the $n$-th R\'enyi entropy.
In particular, we compute a partition function of a $n$-sheeted Riemann surface \cite{cc-09}, with a branch-cut starting at the interface and extending along the half-system. This approach requires the expectation value of the twist fields inserted in a bounded geometry, which is directly related to the dynamics of the R\'enyi entropy.
We provide a general derivation of the linear growth of the entanglement entropy which does not refer to any specific model, and it is based on conformal symmetry only.
Before proceeding with the calculations, it is worth to emphasize the main assumptions behind this approach.
\begin{itemize}
\item The post-quench Hamiltonian is described by the same CFT on both sides of the defect, thus every excitation propagates at the same speed (set to $1$).
\item The interface is scale-invariant, thus no dependence on the incoming momenta is present for the scattering properties across the interface (as explained in Ref. \cite{bbdo-02}).
\item The initial state is a regularized boundary state $\ket{\Psi_0} \sim \exp\left( -\frac{\beta}{4} H\right) \ket{B}$, which is a short-range entangled state sharing features with a thermal state at inverse temperature $\beta$ \cite{cc-16}.
\end{itemize}

\subsection{Bulk and boundary twist field}

Here we introduce and discuss the properties of the twist fields in a defect geometry, stressing the distinction between bulk and boundary fields. First, we briefly review the standard definition of these fields in the absence of interfaces and their connection with entanglement entropy \cite{cc-09}. We consider a 1+1D quantum field theory replicated $n$ times,
and a pair of twist fields $\mathcal{T}(x),\tilde{\mathcal{T}}(x)$, which introduce a branch-cut along the half-line $[x,\infty)$, connecting the $k$-th and the $k\pm 1$-th replica, respectively. A precise definition can be provided via commutation relations between the twist fields and the local fields. In particular, one requires that \cite{ccd-08,bd-16}
\be
\mathcal{T}(x) \mathcal{O}_{k}(y) = \begin{cases} \mathcal{O}_{k+1}(y)\mathcal{T}(x), \quad x<y\\ \mathcal{O}_{k}(y)\mathcal{T}(x), \quad \text{otherwise}, \end{cases}, \qquad \tilde{\mathcal{T}}(x) \mathcal{O}_{k}(y) = \begin{cases} \mathcal{O}_{k-1}(y)\tilde{\mathcal{T}}(x), \quad x<y\\ \mathcal{O}_{k}(y)\tilde{\mathcal{T}}(x), \quad \text{otherwise}, \end{cases}
\label{eq:twist_def}
\eeq
for any local operator $\mathcal{O}_k$ inserted at a point of the $k$-th replica (here $k=1,\dots,n$ is a replica index and its values are identified up to $k=k+n$). In a CFT, the twist fields are primary fields with scaling dimension $\Delta$ given by \cite{cc-09}
\be
\Delta = \frac{c}{12}\left( n- \frac{1}{n}\right),
\label{eq:Delta_twist}
\eeq
and $c$ is the central charge of the theory. Their two-point function in the (homogeneous) ground-state is thus fixed by scaling symmetry only \cite{DiFrancesco-97}, and reads
\be
\la \mathcal{T}(x_1)\tilde{\mathcal{T}}(x_2)\ra = \frac{1}{|x_1-x_2|^{2\Delta}}.
\eeq

The connection between the twist fields and entanglement arises as follows. One considers the ground-state $\ket{\Psi_0}$ of a CFT, and constructs the reduced density matrix $\rho_A$ associated to a subsystem $A$ \cite{ccd-09}. For an interval $A=[0,\ell]$, one can show that for any integer $n\geq 1$ \cite{cc-09}
\be
\text{Tr}(\rho^n_A) = \epsilon^{2\Delta}\la \mathcal{T}(0)\tilde{\mathcal{T}}(\ell) \ra, 
\eeq
with $\epsilon$ being a UV-cutoff. Putting these informations together, one can eventually express the $n$-th R\'enyi entropy of $A$ as \cite{cc-09}
\be
S_n = \frac{1}{1-n} \log \text{Tr}(\rho^n_A) = \frac{c}{6}\left( 1+\frac{1}{n}\right)\log \frac{\ell}{\epsilon},
\label{eq:Sn_tr_inv}
\eeq
which gives the celebrated relation \cite{hlw-94}
\be
S\equiv S_1 = \frac{c}{3}\log \frac{\ell}{\epsilon}
\eeq
for the von Neumann entropy, once the analytical continuation $n\rightarrow 1$ is taken.

While the result \eqref{eq:Sn_tr_inv} is specific to the translational invariant ground-state, the definition \eqref{eq:twist_def} and the (bulk) scaling dimension \eqref{eq:Delta_twist} are general. However, there is an important caveat that arises for a bounded geometry, say an interval $[0,L]$. The value of the scaling dimension \eqref{eq:twist_def} refers only to a twist field $\mathcal{T}(x)$ inserted at a bulk point ($0<x<L$), as it is well known that boundary effects do not affect bulk dimensions \cite{DiFrancesco-97}. Nevertheless, this is no longer true if $x$ is a boundary point ($x=0,L$). In that case, the dimension of the twist field, which is now a boundary field, takes in general another value, dubbed as boundary scaling dimension. Indeed, if we consider the subsystem $A = [0,\ell]$, one finds for $\ell \ll L$ \cite{cd-08}
\be
S_n \simeq \frac{1}{1-n}\log \la \mathcal{T}(0)\tilde{\mathcal{T}}(\ell) \ra  = \frac{c}{12}\left(1+\frac{1}{n}\right)\log \frac{\ell}{\epsilon}+ \dots
\label{eq:Sn_bound}
\eeq
a relation which suggests immediately that the dimension of the boundary twist field $\mathcal{T}(0)$ vanishes. This is expected on physical grounds, since the only contribution to the entanglement is given by the correlations localized around the point $x=\ell$.

So far we have considered a single theory, with or without boundaries, and we have discussed the entanglement of a single interval. Now, we generalize the construction above for a system made by two CFTs joined together through an interface. The two CFTs are denoted by $\text{CFT}_1$ and $\text{CFT}_2$, assumed here to be two copies of the same theory, and they extend over the regions $x \in [0,L]$ and $[-L,0]$ respectively. An interface is inserted at $x=0$, which is a defect line extended over the Euclidean time, and gives a field theoretical representation of the impurity (see Refs. \cite{ss-08,bb-15}). We also need to specify the boundary conditions at $x=\pm L$. Here, we only assume they do not mix the two theories, so that their precise features are essentially irrelevant for the entanglement between the two halves. We call \textit{unfolded picture} the geometry described so far. Via the so-called folding procedure, it is possible to describe the same system as the theory $\text{CFT}_1\otimes \text{CFT}_2$ extended over the region $x\in [0,L]$, a description which takes the name of \textit{folded picture} (see Ref. \cite{gm-17} for further details). Comparing the two pictures, every point $x\in [0,L]$ in the folded geometry corresponds to a pair of points $x,-x$ (associated to $\text{CFT}_1$ and $\text{CFT}_2$ respectively) in the unfolded one. Moreover, the interface is now described via the boundary condition (BC) at $x=0$, denoted by $b$, and similarly the BC at $x=L$ in the folded picture, denoted by $b'$. Note that the latter is just the product of the BCs at $x=\pm L$ in the unfolded picture.

\begin{figure}[thb]
\center
\includegraphics[width=0.7\textwidth]{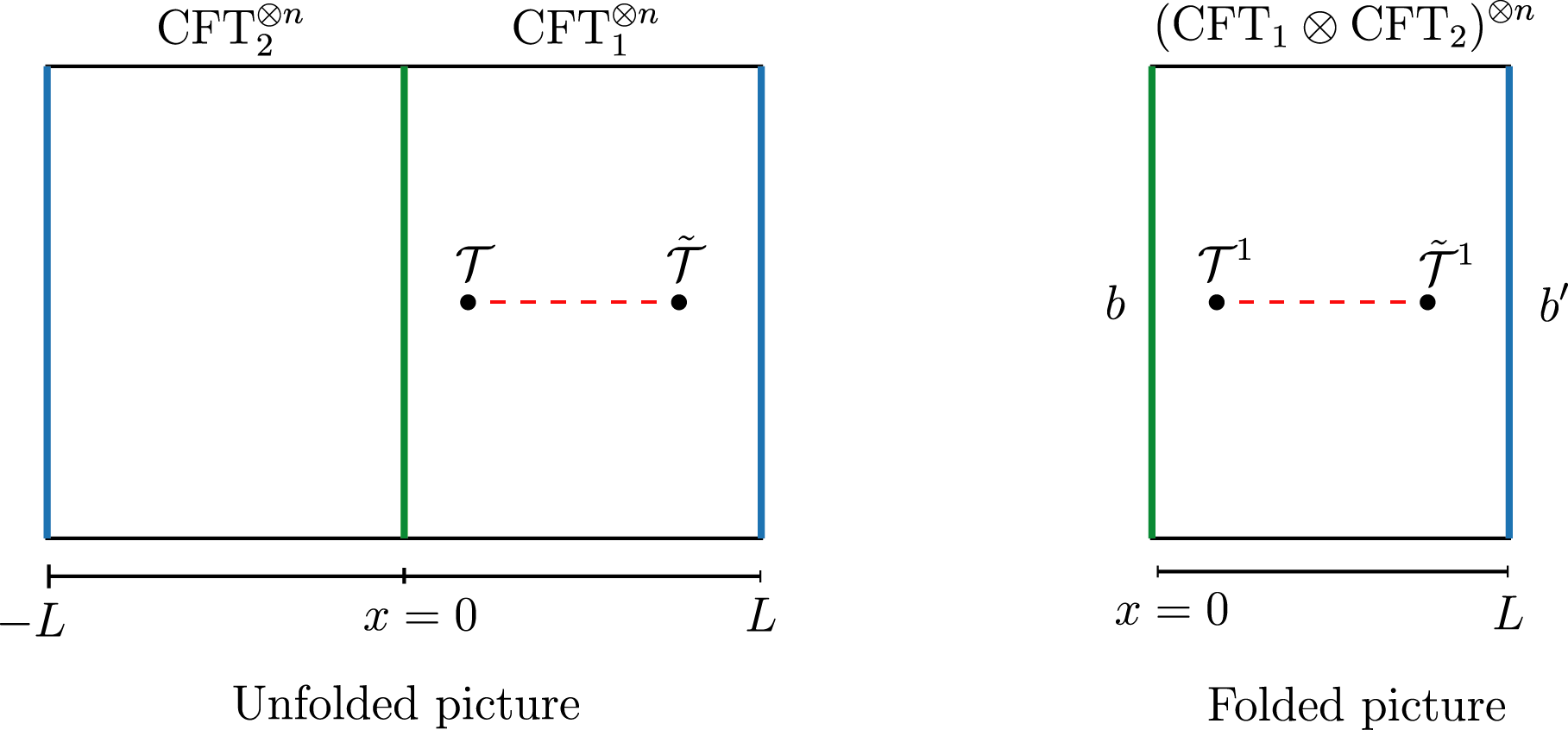}
\caption{Representation of a system made by two CFTs joined through an interface in the unfolded (left) and folded (right) picture.  The interface and the open boundaries are shown by green and blue lines, 
and are associated to boundary conditions of type $b$ and $b'$, respectively.}
\label{fig:Fold_unfold}
\end{figure}

The last ingredient we need to establish the correspondence among the two pictures is the characterization of the branch-cuts in the folded geometry. The idea is to introduce a pair of twist fields $\mathcal{T}^j,\tilde{\mathcal{T}}^j$ ($j=1,2$) which act nontrivially on the degrees of freedom of $\text{CFT}_{1}$ or $\text{CFT}_{2}$ only, and use them as the building blocks for the entanglement measures among the two theories. We propose a definition, which is nothing but a generalization of Eq. \eqref{eq:twist_def}, as follows
\be
\begin{split}
\mathcal{T}^{j}(x) \mathcal{O}_{j',k}(y) = \begin{cases} \mathcal{O}_{j',k+1}(y)\mathcal{T}^{j}(x), \quad x<y,\quad j=j'\\ \mathcal{O}_{j',k}(y)\mathcal{T}^{j}(x), \quad \text{otherwise}, \end{cases} \\
\tilde{\mathcal{T}}^{j}(x) \mathcal{O}_{j',k}(y) = \begin{cases} \mathcal{O}_{j',k-1}(y)\tilde{\mathcal{T}}^{j}(x), \quad x<y,\quad j=j'\\ \mathcal{O}_{j',k}(y)\tilde{\mathcal{T}}^{j}(x), \quad \text{otherwise}, \end{cases}
\end{split}
\eeq
with $\mathcal{O}_{j,k}(y)$ being local operators of $(\text{CFT}_1\otimes \text{CFT}_2)^{\otimes n}$ with species index $j=1,2$ and replica index $k=1,\dots,n$. As an illustrative example, the insertion of $\mathcal{T}^1(x_1)\tilde{\mathcal{T}}^1(x_2)$ at two points $0<x_1<x_2<L$ of the folded geometry, corresponds to the insertion of $\mathcal{T}(x_1)\tilde{\mathcal{T}}(x_2)$ in the unfolded picture, which is a branch-cut along a segment of $\text{CFT}_1$. Similarly, $\mathcal{T}^2(x_1)\tilde{\mathcal{T}}^2(x_2)$ introduces a branch-cut on $[-x_2,-x_1]$ in the unfolded picture.
We summarize the construction above in Fig. \ref{fig:Fold_unfold} which provides a pictorial representations of the insertion of the branch-cuts in the folded/unfolded pictures.

As long as the twist fields $\mathcal{T}^j$ are inserted at bulk points, their (bulk) scaling dimension is simply given by Eq. \eqref{eq:Delta_twist}. In contrast, the dimension of $\mathcal{T}^j(L)$ is expected to be zero, since the two theories are not coupled at $x=L$ and one can use the results available for a single theory. Moreover, we argue that the scaling dimension of $\mathcal{T}^j(0)$ is nontrivial, and strongly depends on the boundary condition $b$ (and $n$), representing the interface: we denote it by $\Delta_b$. We thus compute the R\'enyi entropy among the two halves as
\be
S_n = \frac{1}{1-n}\log \left(\epsilon^{\Delta_b}\la \mathcal{T}^1(0)\tilde{\mathcal{T}}^{1}(L) \ra \right) = \frac{\Delta_b}{n-1}\log \frac{L}{\epsilon} + \dots,
\label{eq:Sn_def}
\eeq
where the last equality follows from scaling arguments only, and subleading terms of order $O(1)$ have been neglected. More precisely, if one applies a scale transformation, say $L \rightarrow L \lambda$, then the geometry becomes $[0,\lambda L]$ and the twist fields are inserted at the boundary of this new geometry ($x=0,\lambda L$). An additional scale factor $\lambda^{\Delta_b}$ appears due to the scaling transformation of the twist fields. Taking for example $\lambda = 1/L$ for the scaling parameter, one completely fixes the $L$-dependence of the two-point function which gives directly Eq. \eqref{eq:Sn_def}.

Eventually, one concludes that the entropy grows as the logarithm of the size, with an  interface dependent prefactor, a result which was firstly obtained for the free boson in \cite{ss-08} and for Ising/free fermions in \cite{bb-15}. In particular, in the limit $n\rightarrow 1$ one has
\be
S = \frac{\ceff}{6}\log \frac{L}{\epsilon},
\eeq 
where $\ceff$ is a parameter dubbed as \textit{effective central charge}. Comparing the last expression with our formula \eqref{eq:Sn_def}, we establish the equivalence
\be
\frac{\ceff}{6} = \underset{n\rightarrow 1}{\lim} \frac{\Delta_b}{n-1}.
\label{eq:ceff_equiv}
\eeq
It tells us that the origin of the nontrivial scaling of entanglement for systems with conformal defects can be traced back to a nontrivial boundary scaling dimension of the twist field $\mathcal{T}^j$.

\subsection{Global quench}

We continue with the study of a global quench for a CFT in the presence of a defect, generalizing the approach of \cite{cc-16} valid without the defect. We mention that a similar method to tackle the problem has been already employed in \cite{wwr-17}. Here, we consider instead a novel technique based on the use of the boundary twist fields.

\begin{figure}[thb]
\center
\includegraphics[width=0.7\textwidth]{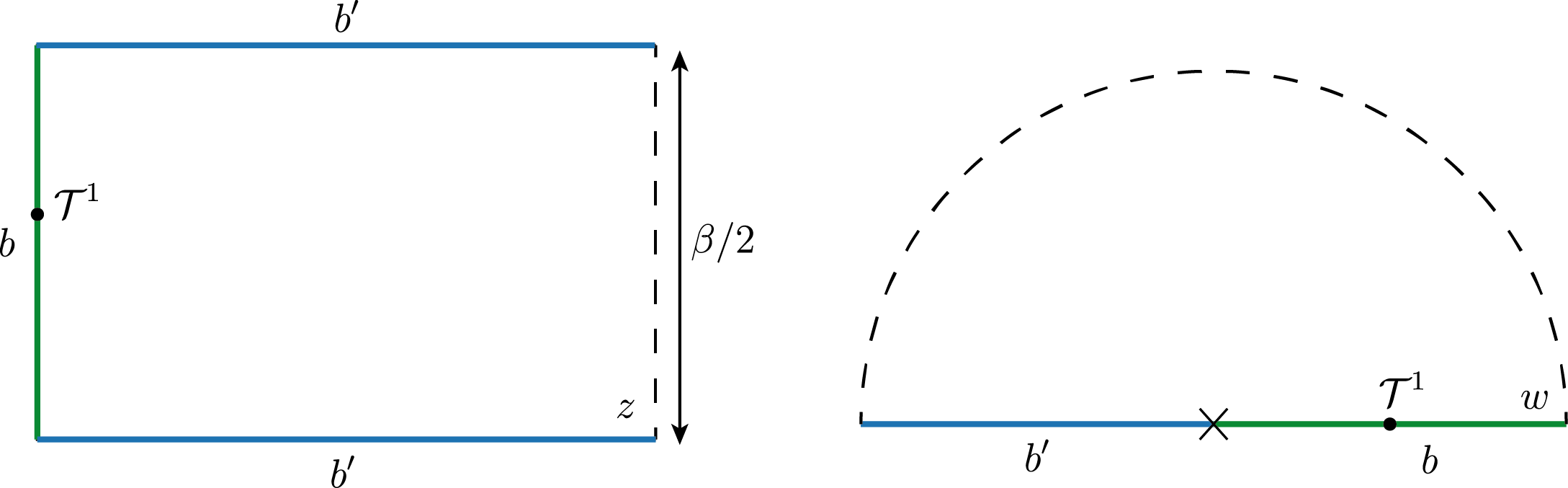}
\caption{Euclidean boundary geometry describing the global quench in the $z$ (left) and $w$ (right) variable. Green/blue colors denote boundary conditions of type $b/b'$, respectively.}
\label{fig:Glob_quench}
\end{figure}

We work in the folded picture, introduced in the previous subsection, and consider the theory $\text{CFT}_1\otimes \text{CFT}_2$ extended over the spacial region $[0,L]$. We take the initial state
\be
\ket{\Psi_0} \sim \exp\left( -\frac{\beta}{4} H\right) \ket{b'},
\eeq
with $\ket{b'}$ being a boundary state associated to the BC $b'$, $H$ the post-quench Hamiltonian and $\beta$ a parameter representing the inverse temperature of the initial state. Within this choice, the initial state is short-range entangled, having a finite correlation length of order $\beta$. Since $\ket{\Psi_0}$ is not an eigenvector of $H$, the unitary dynamics induced by the Hamiltonian is nontrivial. We consider time scales much shorter than the size $L$, $t\ll L$, so that finite-size effects are negligible. We express the R\'enyi entropy between the two halves as
\be
S_{n}(t)= \frac{1}{1-n}\log\left(  {}^{\otimes  n}\bra{\Psi_0}e^{iHt} \mathcal{T}^1(0)  \tilde{\mathcal{T}}^1(L) e^{-iHt}\ket{\Psi_0}^{\otimes n}\right),
\label{eq:Glob_quench_def0}
\eeq
with $\ket{\Psi_0}^{\otimes n} = \ket{\Psi_0}\otimes \dots \otimes \ket{\Psi_0}$ being a state of the $n$-replica theory $(\text{CFT}_1 \otimes \text{CFT}_2)^{\otimes n}$. Since $\tilde{\mathcal{T}}^1(L)$ has dimension $0$, we simply drop it from Eq. \eqref{eq:Glob_quench_def0}, a procedure which is physically motivated by the fact that its insertion should not affect the entanglement among the two halves (see \cite{ccd-08} for futher details). Taking into account the correct normalization of the  state, we finally end up with
\be
S_n(t)= \frac{1}{1-n}\log  \frac{  {}^{\otimes n}\bra{b'}e^{-H(\beta/4-it)} \mathcal{T}^1(0) e^{-H(\beta/4+it)}\ket{b'}^{\otimes n}}{ {}^{ \otimes n}\bra{b'}e^{-H\beta/2} \ket{b'}^{\otimes n}}
\label{eq:Glob_quench_def}.
\eeq
We now proceed further with the evaluation of the expectation value appearing in Eq. \eqref{eq:Glob_quench_def}. We work in the Euclidean theory, described as the bounded geometry (strip)
\be
\text{Re}(z)\in [0,+\infty), \quad \text{Im}(z) \in [-\beta/4,\beta/4],
\label{eq:Strip_z}
\eeq
obtained in the limit $L\rightarrow \infty$. We insert the twist field $\mathcal{T}^1$ at the position $z=i\tau$ of the strip, with $\tau$ being the Wick rotated time
\be
\tau = it.
\eeq
To compute $\la \mathcal{T}^1(z=i\tau)\ra$, we map the strip \eqref{eq:Strip_z} onto the upper half-plane $\text{Im}(w)\geq 0$ through a sequence of transformations $z\rightarrow \zeta \rightarrow w$ defined by
\be
\zeta(z) = \sin \frac{2\pi i z}{\beta}, \quad w(\zeta) = \frac{1+\zeta}{2(1-\zeta)}.
\eeq
Via this change of variables, the corners at $z=\pm i\beta/4$ are mapped onto the points $w=0,\infty$ respectively. In this new geometry a change of boundary conditions, from type $b'$ to type $b$, appears at $w=0$ along the real line $\text{Im}(w)=0$. In Fig. \ref{fig:Glob_quench} we represent the insertion of the twist field both in the strip geometry ($z$ variable) and the upper half-plane ($w$ variable).

At this point, we observe explicitly that the depicted geometry (upper-half plane) shows a scale-symmetry
\be
w \rightarrow \lambda w, \quad \lambda >0.
\eeq
Indeed, the boundary conditions $b',b$ are scale-invariant and the point $w=0$, representing the change of BCs, is kept fixed. We use this property to infer the value of the one-point function along the  boundary of type $b$ ($\text{Im}(w)=0, w>0$) which is
\be
\la \mathcal{T}^1(w)\ra \propto \frac{1}{|w|^{\Delta_b}},
\eeq
up to a $w$-independent proportionality constant not further specified.

We now go back to the original geometry, employing the transformation law of the twist field $\mathcal{T}^1$, which is a primary operator. We firstly apply $w \rightarrow \zeta$, reaching to
\be
\la \mathcal{T}^1(\zeta)\ra = \left|\frac{dw}{d\zeta}\right|^{\Delta_b}\la \mathcal{T}^1(w)\ra \propto \left| \frac{1}{(1-\zeta)^2}\right|^{\Delta_b} \left|\frac{1-\zeta}{1+\zeta}\right|^{\Delta_b} = \left|\frac{1}{1-\zeta^2}\right|^{\Delta_b},
\eeq
and similarly from the map $\zeta \rightarrow z$ we get
\be
\la \mathcal{T}^1(z)\ra = \left|\frac{d\zeta}{dz}\right|^{\Delta_b}\la \mathcal{T}^1(\zeta)\ra \propto \left| \frac{4\pi}{\beta \cos( 2\pi i z/\beta) } \right|^{\Delta_b}.
\eeq
Performing the Wick rotation $\tau = it$, for large values $t/\beta$ we obtain
\be
\la \mathcal{T}(z=i\tau)\ra \sim \left| \frac{8\pi}{\beta} e^{-2\pi t/\beta} \right|^{\Delta_b},
\eeq
which leads directly to
\be
S_{n}(t)-S_n(0) = \frac{1}{1-n}\log \frac{\la \mathcal{T}^1(z=i\tau)\ra}{\la \mathcal{T}^1(z=0)\ra} \simeq \frac{\Delta_b}{n-1} \frac{2\pi t}{\beta},
\label{eq:Lin_Growth}
\eeq
the main result of this subsection. One can further take the analytical continuation $n\rightarrow 1$ from Eq. \eqref{eq:Lin_Growth} to extract the entanglement entropy, and, using Eq. \eqref{eq:ceff_equiv}, one gets
\be
S_{1}(t)-S_1(0) \simeq \frac{\pi \ceff t}{3\beta}.
\eeq
We conclude with some remarks regarding the derivation provided above. For instance, even if we did not attempt to calculate explicitly the scaling dimension $\Delta_b$ of the boundary twist field, we got a precise relation between equilibrium \eqref{eq:Sn_def} and non-equilibrium entropy \eqref{eq:Lin_Growth} based on conformal invariance only. Moreover, the same derivation can be applied to the evolution of any boundary (primary) operator, and it is not specific of the twist field. We finally stress again that the linear growth in \eqref{eq:Lin_Growth} is expected to be valid for $\beta \ll t \ll L$, so that the short-time effects, as well as the long-time recurrences, are neglected.

\section{Lattice results\label{sec:lattice}}

In the following we consider a lattice realization of the global quench.
Our main goal is to derive, analogously to the CFT calculations, a formula that
relates the entropy growth in the presence of a defect to that in a homogeneous chain.
We show that, in contrast to CFT results that predict the appearance of the same 
proportionality factor between the R\'enyi entropies as in the ground state,
this is not true in general for the lattice. We first introduce our model
and summarize the known results about its ground-state entanglement,
followed by a discussion of the quench setup.

\subsection{Model and setup\label{sec:model}}

We consider a fermionic hopping chain of length $2L$ with Hamiltonian
\eq{
\hat H = \sum_{m,n=-L+1}^{L} H'_{m,n} c_m^\dag c_n ,
\label{Hdef}}
where $c^\dag_m$ and $c_m$ are fermionic creation/annihilation operators satisfying
anticommutation relations $\{ c^\dag_m, c_n \} = \delta_{m,n}$. We focus on a particular
form of a defect located in the center of the chain, with the nonzero elements of the
hopping matrix given by
\eq{
H'_{m,m+1}= H'_{m+1,m} = \left\{
\begin{array}{ll}
-1/2 & \quad m \ne 0 \\
-\lambda/2  & \quad m = 0
\end{array}\right. , \qquad
H'_{0,0} = -H'_{1,1}=\sqrt{1-\lambda^2}/2 \, .
\label{confdef}}
The defect of strength $\lambda$ thus consists of a single modified hopping amplitude
as well as local chemical potentials across sites $0$ and $1$. Throughout the manuscript,
we use the prime notation to refer to various quantities of the defect problem,
whereas the same symbols without prime shall refer to the homogeneous ($\lambda=1$) case.

The defect problem defined by the hopping matrix $H'$ was studied before in \cite{ep-12}
and was shown to have a particularly simple relation to the homogeneous case $H$,
where the single-particle eigenvalues and eigenvectors are given by
\eq{
\phi_k(m) = \sqrt{\frac{2}{2L+1}}\sin\frac{\pi k m }{2L+1}, \qquad
\omega_k = -\cos \frac{\pi k}{2L+1} \,
\label{phik}
}
for $k = 1,\dots,2L$. Indeed, the corresponding quantities for the defect are simply related via \cite{ep-12}
\eq{
\phi'_k(m) = \left\{
\begin{array}{ll}
\alpha_k \phi_k(m) & \quad m \le 0 \\
\beta_k \phi_k(m) & \quad  m \ge 1
\end{array}\right. , \qquad
\omega'_k = \omega_k
\label{phikdef}
}
where the scaling factors read
\eq{
\alpha_k^2 = 1 + (-1)^k \sqrt{1-\lambda^2}, \qquad
\beta_k^2 = 1 - (-1)^k \sqrt{1-\lambda^2} \, .
\label{ab}
}
In other words, the eigenvectors of $H'$ are only rescaled by a different factor on
the left/right hand side of the defect, while the spectrum remains unchanged.

The relation \eqref{phikdef} has important consequences on the transmission
properties of the defect, which are easier to discuss in the thermodynamic limit
$L \to \infty$ of the chain. One then writes the following plane-wave ansatz for the
eigenvectors $\bar \phi_q(m)$ and $\phi_q(m)$ on the left/right hand side of the defect
\eq{
\bar \phi_q(m)=
\begin{cases}
A_1 \, \ee^{iqm} + B_1 \, \ee^{-iq(m-1)} & q>0 \\
A_2 \, \ee^{iqm} & q<0
\end{cases},
\qquad
\phi_q(m)=
\begin{cases}
A_2 \, \ee^{iqm} & q>0 \\
A_1 \, \ee^{iqm} + B_1 \, \ee^{-iq(m-1)} & q<0
\end{cases}.
\label{philr}}
For a right-moving $(q>0)$ particle, $\bar \phi_q(m)$ is just the superposition
of an incoming and reflected wave, whereas $\phi_q(m)$ corresponds to the
transmitted wave component (and vice-versa for $q<0$). The coefficients
are fixed by matching the solutions at the defect via
\eq{
\lambda \bar \phi_q(0) + \sqrt{1-\lambda^2} \phi_q(1) = \phi_q(0) \, ,
\qquad
\lambda \phi_q(1) - \sqrt{1-\lambda^2} \bar\phi_q(0) = \bar \phi_q(1) \, ,
}
and yield the following transmission and reflection amplitudes
\eq{
s(q) = \frac{A_2}{A_1} = \lambda \, , \qquad
r(q) =  \frac{B_1}{A_1} = -\textrm{sign}(q) \sqrt{1-\lambda^2} \, .
\label{sr}}
The transmission $T=|s(q)|^2=\lambda^2$ and reflection $R=|r(q)|^2=1-T$
coefficients are thus independent of the wavenumber, hence it has been dubbed
as a \emph{conformal defect} \cite{ep-12}. Note, however, that there is a nontrivial
sign factor in the reflection amplitude $r(q)$, required by the orthonormality
of the basis, which also yields $|A_1|^2=1/2\pi$ for the incoming amplitude.

The simple structure \eqref{phikdef} of the eigenvectors has also important
implications on the ground-state entanglement properties between the two halves
$A=\left[-L+1,0\right]$ and $B=\left[1,L\right]$ of the chain. These follow from the fermionic reduced correlation $C'_A$ matrix with elements \cite{pe-09} 
\eq{
C'_{m,n} = \braket{c^\dag_m c_n} = \sum_{k=1}^{L} \phi'_k(m)\phi'_k(n) \, ,
\qquad m,n \in A \, .
\label{CA}}
In particular, the entanglement entropy of the reduced density matrix $\rho_A=\Tr_B \, \rho$
is obtained as
\eq{
S = \sum_{l=1}^{L} s(\zeta'_l) \, ,\qquad
s(x) = -x \ln x - (1-x) \ln (1-x) \, ,
\label{S}}
where $\zeta'_l$ are the eigenvalues of $C'_A$. Using \eqref{phikdef} for the matrix
elements in \eqref{CA}, one finds the exact relation for the ground state of the conformal defect
\eq{
C'_A(1-C'_A) = \lambda^2 \, C_A(1-C_A) \, ,
\label{Crelgs}}
which yields an analogous relation between the eigenvalues. Equivalently, one could
rewrite \eqref{Crelgs} in terms of the single-particle entanglement spectrum
\eq{
\zeta'_l = \frac{1}{\ee^{\varepsilon'_l}+1}\, ,
}
which then yields the relation
\eq{
\cosh \frac{\varepsilon'_l}{2}=\frac{1}{\lambda}\cosh \frac{\varepsilon_l}{2} \, .
\label{omeps}}

The latter relation can then be used to find the  scaling of the entanglement
entropy which reads
\eq{
S = \kappa(\lambda) \ln L \, ,
\label{Sdef}}
where the prefactor of the leading logarithmic term is obtained as \cite{ep-10}
\eq{
\kappa(s)=-\frac{1}{\pi^2} \bigg\{ \Big[ (1+s) \ln (1+s) + (1-s) \ln (1-s) \Big] \ln s
+ (1+s) \mathrm{Li_2}(-s) + (1-s) \mathrm{Li_2}(s) \bigg\} \, ,
\label{ceff}}
in terms of the transmission amplitude $s=\lambda$. Note that the prefactor
could also be written as $\kappa(\lambda)=\ceff/6$ via an effective
central charge, which varies smoothly between one and zero, corresponding
to the purely transmissive ($\lambda=1$) and reflective cases ($\lambda=0$).
In fact, \eqref{omeps} can also be used to extract the scaling of the
R\'enyi entropies of integer index $n>1$
\eq{
S_n=\sum_{l=1}^{L} s_n(\zeta'_l) \, ,\qquad
s_n(x) = \frac{1}{1-n} \ln \left[x^n + (1-x)^n \right]  \, .
\label{Sn}}
The result is completely analogous to \eqref{Sdef}, with the corresponding prefactor
given by \cite{pe-12}
\eq{
\kappa_n(\lambda)=\frac{1}{\pi^2}\frac{1}{n-1} \sum_{p=-(n-1)/2}^{(n-1)/2}
\arcsin^2 \left[\lambda \, \sin \left(\frac{p\pi}{n} \right)\right] \, ,
\label{kappan}}
where the sum over $p$ runs over half-integer/integer values for $n$ even/odd.

We now turn our attention to the quench setup.
In order to mimic the global quench scenario of the CFT setting, we choose a gapped
initial Hamiltonian $\hat H_0$ and prepare the chain in its ground state. The simplest
way to open a gap is to add a staggered chemical potential to the Hamiltonian \eqref{Hdef}
of the conformal defect
\eq{
\hat H_0 = \hat H + \mu \sum_{n=-L+1}^{L} (-1)^n c^\dag_n c_n \, .
\label{H0}}
The quench then consists of switching off the potential at time $t=0$ and time evolve
the initial state with $\hat H$.
The size of the gap in $\hat H_0$ is controlled by the mass parameter $\mu$, and 
in the limit $\mu \to \infty$ one obtains the N\'eel state, where each odd lattice site
is occupied by a particle with the even sites being empty.
In fact, the very same initial Hamiltonian \eqref{H0} was considered in Ref. \cite{wwr-17}
to test the CFT predictions for the global quench numerically, finding a good agreement.
Here we revisit this problem more carefully, providing analytic results for the quench
on the lattice that show some deviations from the CFT result.
We start by discussing the homogeneous problem first.

\subsection{Quasiparticle ansatz for homogeneous case\label{sec:quenchhom}}

\begin{figure}[h]
\center
\includegraphics[width=0.6\textwidth]{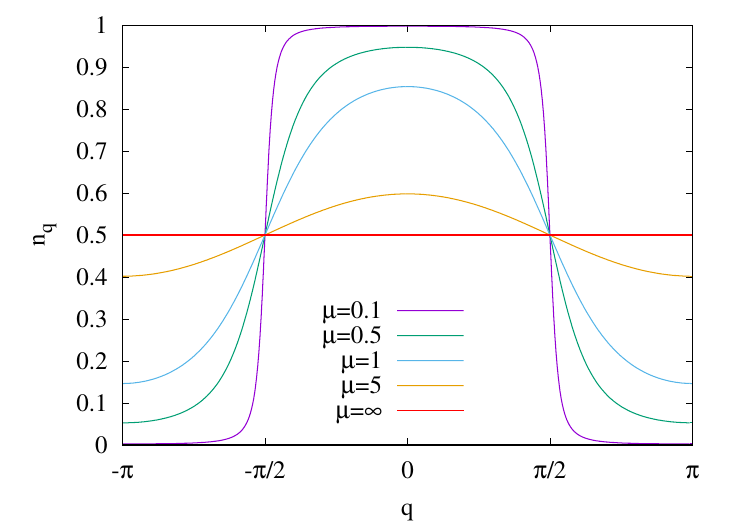}
\caption{Occupation numbers \eqref{nqmu} for various values of $\mu$.}
\label{fig:nqmu}
\end{figure}

The entropy growth after a global quench in an integrable translational invariant chain
can be understood in terms of a quasiparticle picture \cite{cc-05,fc-08,ac-16}. Namely, one assumes that
the initial state is a source of quasiparticle pair excitations that propagate linearly under
the dynamics, and carry the initial short range entanglement to large distances.
In a free-fermion chain, each mode with momentum $q$ propagates with the corresponding
group velocity $v_q=\sin q$, and a pair contributes to the entanglement at time $t$
if one particle is located in subsystem $A$ while the other one in $B$.
The contribution of a given pair is simply the thermodynamic entropy density taken
in the steady state of the dynamics, which is uniquely determined by the occupation numbers
$n_q=\braket{c^\dag_qc_q}$, that are constants of motion. Putting everything together,
in the limit $L \to \infty$ of a semi-infinite subsystem one arrives at the expression
\eq{
S_n(t)= t \int_{-\pi}^{\pi} \frac{\dd q}{2\pi} |v_q| \, s_n(n_q) \, ,
\label{Snqp}}
where the entropy density $s_n$ was defined in \eqref{Sn}, see also \eqref{S} for $n=1$.
The quasiparticle ansatz \eqref{Snqp} in the half-chain geometry thus gives a purely linear growth of
entanglement, $S_n(t)=\gamma_n(\mu,1) t$, where the second argument of the slope refers to $\lambda=1$.
In order to find the slope $\gamma_n(\mu,1)$ as a function
of the quench parameter $\mu$, one has to
evaluate the occupation numbers $n_q$ characterising the steady state after the quench. These are fixed
by the ground state of the staggered chain, and can be found by diagonalizing $\hat H_0$ as shown
in appendix \ref{app:stag}, with the result
\eq{
n_q = \frac{1}{2}\left(1+\frac{\cos q}{\sqrt{\cos^2 q+\mu^2}}\right).
\label{nqmu}}
The occupation function is shown in Fig.~\ref{fig:nqmu} for various values of $\mu$.
In the limit $\mu \to 0$ it converges towards the Fermi sea occupation, whereas for
the N\'eel state, $\mu \to \infty$, it becomes completely flat, $n_q=1/2$.

\begin{figure}[h]
\center
\includegraphics[width=0.49\textwidth]{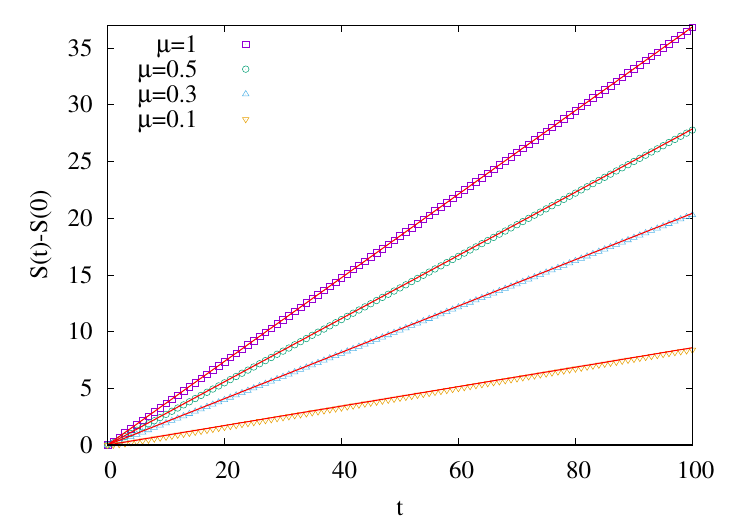}
\includegraphics[width=0.49\textwidth]{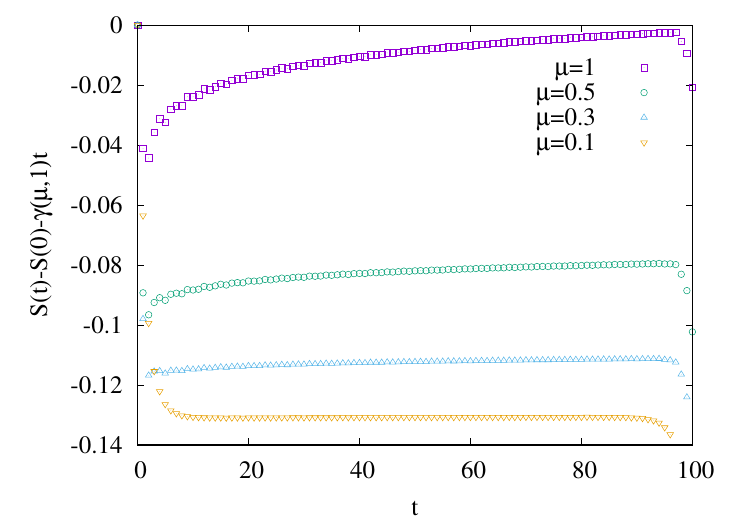}
\caption{Left: entropy evolution for a homogeneous chain ($\lambda=1$) with $L=100$
after a global quench for various values of $\mu$. The red lines show the slopes calculated from \eqref{Snqp} with the occupation \eqref{nqmu}.
Right: deviation from the quasiparticle ansatz.}
\label{fig:entmu}
\end{figure}

With the result \eqref{nqmu} for the occupation at hand, one can now compare
the quasiparticle ansatz \eqref{Snqp} to the entropy obtained from the lattice calculation.
We shall restrict ourselves to the case $n=1$. This can be obtained from \eqref{S},
using the eigenvalues $\zeta_l(t)$ of the time-evolved reduced correlation
matrix $C_A(t)$. The correlations at time $t$ are given by $C(t)=U^\dag C(0)U$,
where $C(0)$ describes the correlations of the staggered ground state, and the
propagator is given by $U=\ee^{-iHt}$. For our numerics we choose $L=100$,
and consider the short-time regime $t<L$, where the presence of the boundaries
cannot yet influence the behaviour.
The result is shown in Fig.~\ref{fig:entmu} for various values of $\mu$.
On the left, the linear growth with slopes $\gamma(\mu,1)\equiv\gamma_1(\mu,1)$
calculated from \eqref{Snqp} are plotted by red solid lines, showing a very good agreement
with the data. Note that we have subtracted the initial value $S(0)$ of the entropy.
On the right of Fig.~\ref{fig:entmu} we show the corrections to the quasiparticle ansatz
by subtracting the linear piece. Interestingly, the leading correction seems to be logarithmic
in time, albeit with a very small prefactor that increases for larger $\mu$.

\subsection{Quench with a defect: short times\label{sec:quenchdef}}

We now move on to consider the quench with the defect. Motivated by the CFT results,
our goal is to establish a relation between the defect and the homogeneous cases,
analogous to the one \eqref{Crelgs} found for the ground state.
The first step is to calculate the defect propagator $U'=\ee^{-iH't}$.  Using \eqref{philr},
the matrix elements on the same side of the defect are obtained as
\eq{
U'_{m,n}=
\int_{-\pi}^{\pi} \frac{\dd q}{2\pi} \, \ee^{-i\omega_q t} 
\left[\ee^{-iq(m-n)} \pm \sqrt{1-\lambda^2} \, \ee^{-iq(m+n-1)} \right]=
U_{m,n} \pm \sqrt{1-\lambda^2} \, U_{m,1-n},
\label{U1}}
where the $\pm$ sign refers to the case $m,n \ge 1$ and $m,n\le 0$, respectively.
The defect propagator thus differs from the homogeneous one in the presence of a
reflected component that is multiplied by the corresponding amplitude.
On the other hand, for the off-diagonal terms with $m \le 0$ and $n \ge 1$ as well as
$m \ge 1$ and $n \le 0$ one obtains
\eq{
U'_{m,n}=
\lambda \int_{-\pi}^{\pi} \frac{\dd q}{2\pi} \, \ee^{-i\omega_q t} \ee^{-iq(m-n)} =
\lambda \, U_{m,n} \, .
\label{U2}}
Hence, for indices on opposite sides of the defect the propagator is just multiplied
by the transmission amplitude.

For the calculation of the entropy we need the time-evolved matrix
$C'(t)=U'^\dag C'(0)U'$, where $C'(0)$ contains the ground-state correlations of the
staggered Hamiltonian \eqref{H0} with a defect. In general, this is a rather complicated
object of which we do not have a closed-form expression. Hence, for the purpose of
this calculation we shall restrict ourselves to the N\'eel case ($\mu \to \infty$), where
$C'(0)$ is diagonal and independent of $\lambda$. In order to establish the analogue
of \eqref{Crelgs}, let us first note that, due to the purity of the time evolved state,
the full correlation matrix satisfies $C'(t)(1-C'(t))=0$. This yields
\eq{
C'_A(t)(1-C'_A(t)) = C'_{AB}(t)C'_{BA}(t) \, ,
\label{CACAB}}
where $C'_{AB}(t)$ is the off-diagonal part of the correlation matrix, and
$C'_{BA}(t)=C'^\dag_{AB}(t)$. Its matrix elements with the N\'eel initial state read
\eq{
C'_{m,n}(t) = \sum_{l=-\infty} ^{\infty}U'^*_{2l-1,m}U'_{2l-1,n} \, ,
}
where $m \le 0$ and $n \ge 1$. Using the relations \eqref{U1} and \eqref{U2}, this leads to
\eq{
\sum_{l \le 0} (U^*_{2l-1,m} - \sqrt{1-\lambda^2} \, U^*_{2l-1,1-m}) \lambda U_{2l-1,n}
+\sum_{l \ge 1} \lambda U^*_{2l-1,m} (U_{2l-1,n} + \sqrt{1-\lambda^2} \, U_{2l-1,1-n}) \, ,
}
which, using the reflection symmetry of the propagator $U_{m,n}=U_{1-m,1-n}$, can be rewritten as
\eq{
\lambda \sum_l U^*_{2l-1,m}U_{2l-1,n} + \lambda\sqrt{1-\lambda^2}
\sum_{l \ge 1} ( U^*_{2l-1,m} U_{2l-1,1-n} - U^*_{2l,m} U_{2l,1-n}) \, .
\label{CAB}}
In other words, one finds
\eq{
C'_{AB}(t) = \lambda C_{AB}(t) + \lambda\sqrt{1-\lambda^2} \Delta C_{AB}(t) \, ,
\label{CAB2}}
where $\Delta C_{AB}(t)$ denotes the nonvanishing contribution from the second sum in \eqref{CAB}.

The expression obtained in \eqref{CAB2} is thus not quite the one we expect. Indeed,
the exact analogue of the ground-state relation \eqref{Crelgs} would be reproduced via \eqref{CACAB}
only for $\Delta C_{AB}(t)\equiv 0$, which is not the case here.
The interpretation of this result is as follows. The first term of $\eqref{CAB2}$
describes the process where one of the quasiparticles gets transmitted across the defect,
picking up an amplitude $\lambda$, while its pair simply propagates in the homogeneous chain.
On the other hand, the prefactor $\lambda\sqrt{1-\lambda^2}$ of the second term
shows, that the corresponding contribution is carried by the partially transmitted
and reflected wave. Although such a process generates some nontrivial correlations
between the two halves of the chain, we argue that its contribution to the entanglement
must be subleading. Indeed, the main idea of the quasiparticle picture is that the entanglement
is carried by counterpropagating modes created at time $t=0$ all along the chain.
This contribution gets reduced by the transmission amplitude $s(q)=\lambda$ which
is completely independent of $q$. On the other hand, however, the reflection amplitude
$r(q)$ in \eqref{sr} contains a $\mathrm{sign}(q)$ factor, and thus the dynamics 
generates a distructive interference between the partially reflected contributions of pairs
emitted from opposite sides of the defect. In turn, the nonvanishing of the $\Delta C_{AB}(t)$
term is actually due to the imperfect symmetry of the initial state.

\begin{figure}[htb]
\center
\includegraphics[width=0.49\textwidth]{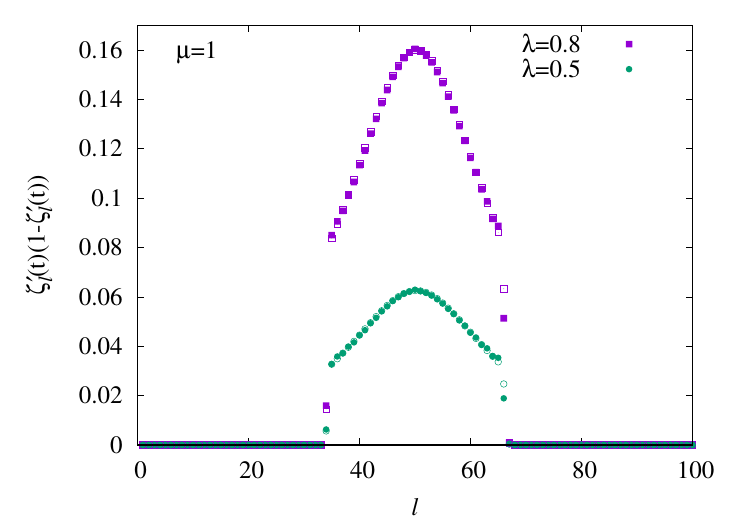}
\includegraphics[width=0.49\textwidth]{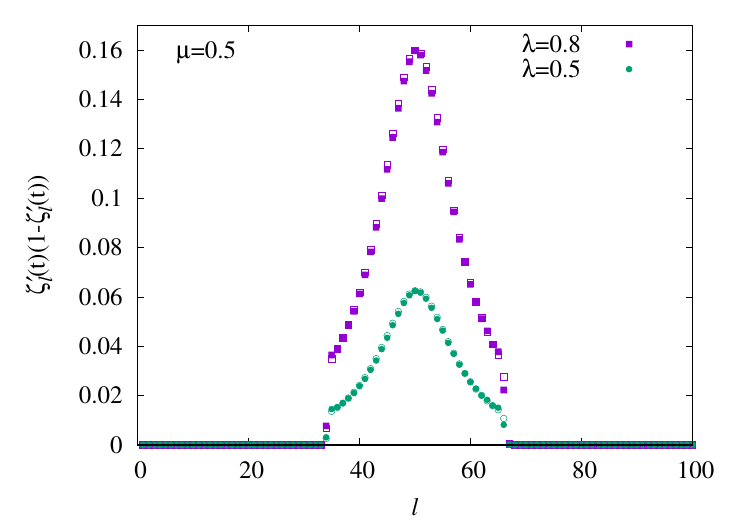}
\caption{Numerical check of the relation \eqref{zetadef}, with the left/right hand side of the equation shown by
the full/empty symbols, for $L=100$, $t=50$ and two different values of $\mu$.}
\label{fig:zetamu}
\end{figure}

The situation for generic $\mu$ is even more complicated, as the initial correlations
become more and more long-ranged as $\mu\to 0$ and the contributions from the off-diagonal
blocks of $C'(0)$ become relevant as well. The corresponding relation \eqref{CAB2}
thus becomes more complicated, with additional terms appearing with prefactors that
mirror the relevant scattering process. Nevertheless, similarly to our above argument,
we expect that these are due to imperfect destructive interference and thus are
higher order effects from the viewpoint of entanglement dynamics.
Hence, we expect that the relation for the eigenvalues
\eq{
\zeta'_l(t)(1-\zeta'_l(t)) \simeq \lambda^2 \zeta_l(t)(1-\zeta_l(t))
\label{zetadef}}
should still hold approximately, even though the corresponding relation
between the matrices $C'_A(t)$ and $C_A(t)$ does not hold, not even in a perturbative sense.
The eigenvalue relation is tested in Fig.~\ref{fig:zetamu} for a chain of size
$L=100$, time $t=50$ and two different values of $\mu$, with full/empty symbols
corresponding to the left/right hand side of Eq. \eqref{zetadef}.
One can see a very good overlap between the two quantities, with only slight shifts
between the symbols, except for the spectral edges where the discrepancy becomes larger.
Similar results can be obtained for arbitrary times $t<L$, with the number of nonzero
data points increasing with $t$.

\begin{figure}[htb]
\center
\includegraphics[width=0.49\textwidth]{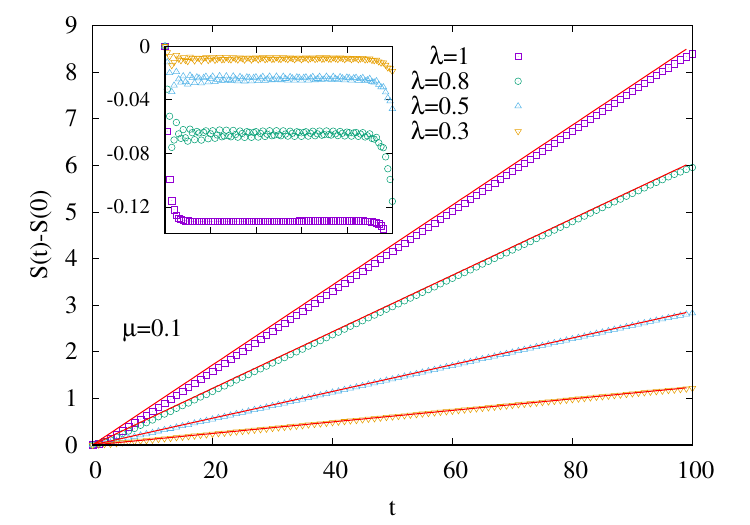}
\includegraphics[width=0.49\textwidth]{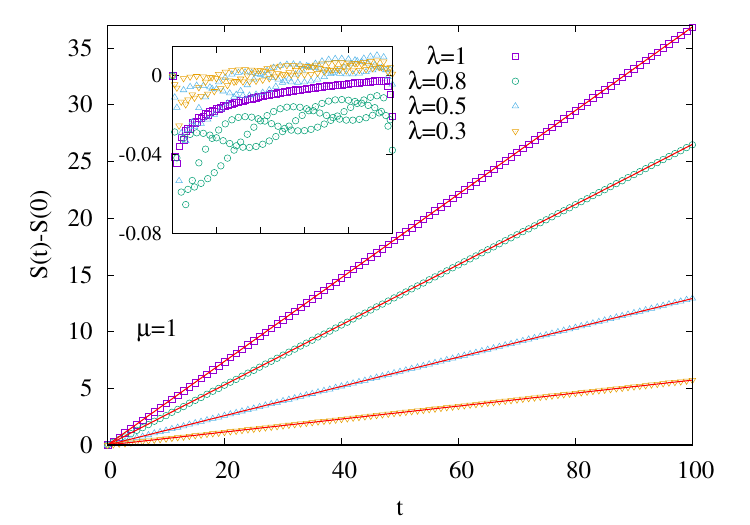}
\caption{Entropy growth after a global quench with mass parameters $\mu=0.1$ (left) and $\mu=1$ (right),
for various defect strengths $\lambda$ and $L=100$. The red lines show the quasiparticle ansatz with
slope \eqref{gammulam}, while the insets show the corresponding deviations.}
\label{fig:dentalpha}
\end{figure}

According to the quasiparticle picture, relation \eqref{zetadef} should actually be
interpreted as a relation between the occupation functions after the quench
\eq{
n'_q(1-n'_q) = \lambda^2 n_q(1-n_q) \, .
\label{nqrel}}
Using the homogeneous result \eqref{nqmu}, this equation can be solved for the defect
occupations as
\eq{
n'_q= \frac{1}{2}\left( 1 \pm \sqrt{1-\lambda^2\frac{\mu^2}{\cos^2 q+\mu^2}} \right) .
\label{nqp}}
Hence, the slope of the entropy growth is given by
\eq{
\gamma_n(\mu,\lambda)=\int_{-\pi}^{\pi} \frac{\dd q}{2\pi} |v_q| \, s_n(n'_q) \, .
\label{gammulam}}
The result \eqref{gammulam} is compared against the lattice data in Fig.~\ref{fig:dentalpha}
for a small and intermediate value of the mass parameter, and various defect strengths $\lambda$.
One observes a very good agreement, with the deviations from the quasiparticle result shown on the inset.

The subleading term for $\mu = 1$ is likely to be logarithmic in time, with some superimposed oscillations, whereas it seems to be given by a constant for $\mu = 0.1$. However, a closer inspection of the latter case indicates that the corrections are probably still logarithmic, albeit with a tiny prefactor, which is also supported by calculations for $L=200$.

\begin{figure}[htb]
\center
\includegraphics[width=0.49\textwidth]{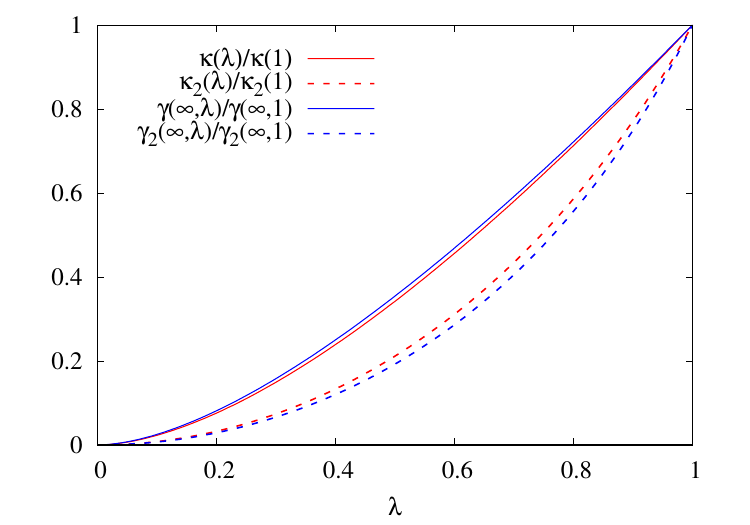}
\includegraphics[width=0.49\textwidth]{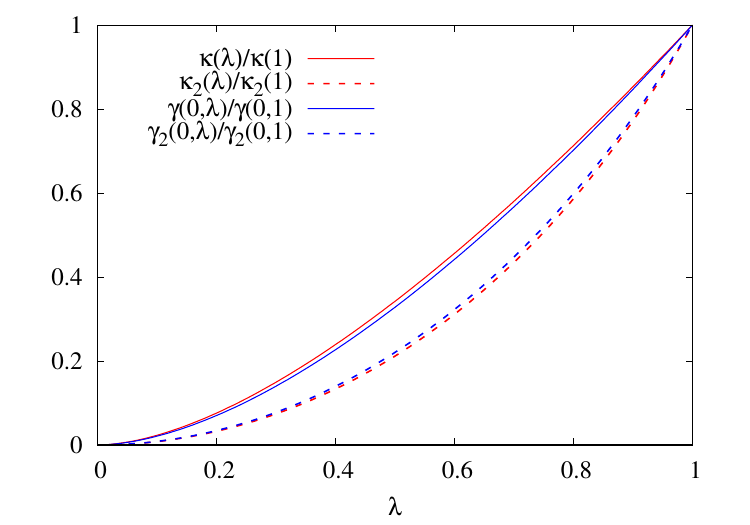}
\caption{Comparison of the slope ratios (blue) to the CFT prediction (red) in the limits
$\mu\to\infty$ (left) and $\mu \to 0$ (right), as a function of $\lambda$.
The solid/dashed lines correspond to $n=1$ and $n=2$.}
\label{fig:slopevsceff}
\end{figure}

With the expression \eqref{gammulam} for the slope at hand, one can now compare
the result to the CFT prediction. This tells us that the slope ratio $\gamma_n(\mu,\lambda)/\gamma_n(\mu,1)$
should be equal to the ground-state entropy ratio $\kappa_n(\lambda)/\kappa_n(1)$.
Let us first consider the N\'eel limit, $\mu \to \infty$, where the occupation $n'_q$ in \eqref{nqp} becomes piecewise constant,
and the integral \eqref{gammulam} simply evaluates to
\eq{
\gamma_n(\infty,\lambda) = \frac{2}{\pi} s_n \left(\frac{1+\sqrt{1-\lambda^2}}{2} \right) .
\label{Scdef}}
Obviously, this result has a rather different analytical behaviour as compared to the ground-state prefactors \eqref{kappan}.
Remarkably, however, the ratios turn out to be very close to each other, as shown by the comparison on the left of Fig.~\ref{fig:slopevsceff}
for $n=1,2$, although the discrepancy increases with $n$.
In fact, in the limit of large mass $\mu$, one does not expect the CFT result to be exact,
since one obtains contributions from the entire Brillouin zone and thus the role of the lattice dispersion enters.
In contrast, in the limit $\mu \to 0$ only the modes around the Fermi level $q=\pm \pi/2$ contribute
(see Fig.~\ref{fig:nqmu}), and one would expect the CFT description to become exact.
Surprisingly, this turns out not to be the case, as demonstrated on the right of Fig.~\ref{fig:slopevsceff}.
Indeed, in the limit $\mu \to 0$, the slope vanishes linearly with the mass and one can find the closed analytical expression for $n \ne 1$
\eq{
\lim_{\mu \to 0} \frac{\gamma_n(\mu,\lambda)}{\mu}=\frac{1}{n-1}
\sum\limits^{(n-1)/2}_{p=-(n-1)/2} \left[1-\sqrt{1-\lambda^2\sin^2 ( \pi p/n)}\right],
\label{gammu0}}
as shown in appendix \ref{app:mu0}. The ratio $\gamma_n(0,\lambda)/\gamma_n(0,1)$ is thus finite, and the structure of \eqref{gammu0} shows a close resemblance to that in \eqref{kappan}, although the expressions in the sum are different. Nevertheless, the mismatch from the ratio $\kappa_n(\lambda)/\kappa_n(1)$ turns out to be very small again, and the curves now approach the CFT limits from the other side, as compared to the $\mu\to\infty$ case. 

For general $\mu$, the behaviour of the discrepancy
$\gamma(\mu,\lambda)/\gamma(\mu,1)-\kappa(\lambda)/\kappa(1)$ between the slope ratios is shown for some fixed values of $\lambda$.
The deviation remains rather small in the entire regime $0.1 \le \mu \le 5$ shown.
In particular, it decreases for $\lambda \to 0$ and $\lambda \to 1$, while the maximal deviations
are observed around $\lambda \simeq 0.5$, similarly to Fig.~\ref{fig:slopevsceff}.
Interestingly, the curves change sign at around $\mu \simeq 0.3$, although they do not
intersect at the same $\mu$ as it might seem. Thus, in general, the lattice quench ratios are governed by a different function of $\lambda$ from the one in CFT, albeit with such a small discrepancy that could not have been found by fitting the data, without the analytical solution of the problem.

\begin{figure}[htb]
\center
\includegraphics[width=0.6\textwidth]{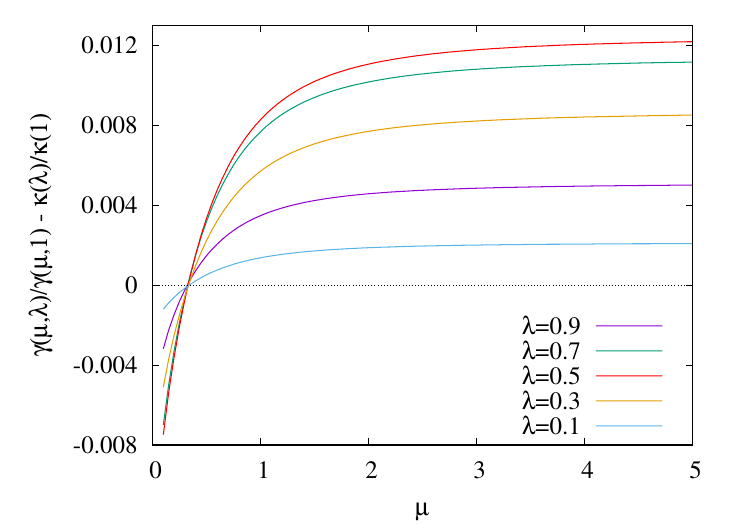}
\caption{Deviation of the slope ratio from the CFT prediction with $n=1$,
plotted in the regime $0.1 \le \mu \le 5$ for various fixed values of $\lambda$.}
\label{fig:slopevsceff2}
\end{figure}

The reason of the discrepancy is that, even in the limit $\mu \to 0$, the occupation $\eqref{nqmu}$ does not correspond to a thermal distribution, which is implicitly assumed in the CFT treatment. In fact, the calculation of the slope can be generalized to this case, since the relation \eqref{nqrel} holds for \emph{arbitrary} initial occupations. Choosing a thermal initial state of the homogeneous chain
\eq{
\bar n_q= \frac{1}{\ee^{\beta \omega_q}+1}
}
with $\omega_q=-\cos q$, the details of the dispersion become irrelevant in the low-temperature limit, $\beta \to \infty$. As outlined in appendix \ref{app:mu0},
the calculation of the corresponding slope $\bar{\gamma}_n(\beta,\lambda)$ can be carried out explicitly by linearizing the dispersion around the Fermi point and yields
\eq{
\lim_{\beta\to\infty} \beta \, \bar \gamma_n(\beta,\lambda)= 2\pi \kappa_n(\lambda) \, .
}
Hence, the slope ratios give exactly the expected CFT result in the limit $\beta\to\infty$.

Finally, one should emphasize that the key connection between the ground-state and
quench scenarios is the eigenvalue relation \eqref{zetadef}, which is indeed identical
for the two cases, even though it is satisfied only approximately for the quench with the
initial state chosen here. It is easy to see that, by choosing instead a reflection symmetric N\'eel
state with only even/odd sites occupied in the left/right half-chain, the second term in
\eqref{CAB} vanishes identically. Hence one has $C'_{AB}=\lambda C_{AB}$ and using
\eqref{CACAB}, the relation becomes exact even on the level of matrix elements.
Furthermore, as shown in appendix \ref{app:corr}, the same holds true for the ground state
of an initial Hamiltonian with a reflection symmetric staggered potential with arbitrary $\mu$.
The derivation relies on a special property of the initial Hamiltonian that is completely
analogous to \eqref{phikdef}, and the relation $C'_{AB}=\lambda C_{AB}$ holds exactly for
finite $L$ and arbitrary times.

\subsection{Entanglement revivals\label{sec:revival}}

\begin{figure}[htb]
\center
\includegraphics[width=0.6\textwidth]{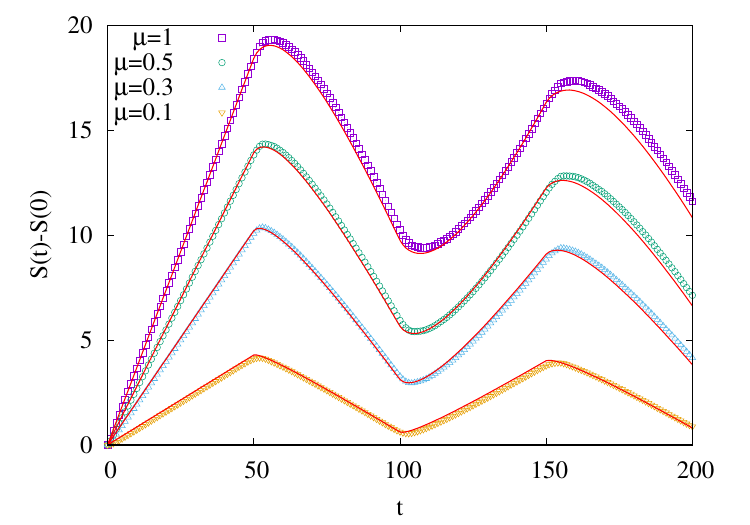}
\caption{Entanglement decay and revival in the homogeneous quench for various $\mu$ and $L=50$.
The red solid lines show the quasiparticle ansatz \eqref{sqprev}.}
\label{fig:entmurev}
\end{figure}

For a chain of finite size, the result found in sec.~\ref{sec:quenchdef} will be modified for times $t>L$, due to particles reflecting from the open boundaries of the chain. Such finite-size effects
result in a sawtooth-like pattern of the entropy, as shown in Fig.~\ref{fig:entmurev} for the
homogeneous case $\lambda=1$.
The decay and revival of the entropy was studied for a periodic chain in \cite{mac-20},
and the result can easily be generalized to the open chain by finding the proper contributions
of the quasiparticle pairs after reflections. Indeed, for a pair with fixed velocity $v_q>0$, the emission
distance measured from the center of the chain must satisfy $x < \min(v_q t-2L(n-1), 2L \, n-v_qt)$ 
in order to contribute after $n$ reflections from the boundary.
Hence, the quasiparticle ansatz reads
\eq{
S_n = \int_{-\pi}^{\pi} \frac{\dd q}{2\pi} \, s_n(n_q) \, 2L
\min \left[ \left\{\frac{|v_q| t}{2L}\right\}, 1-\left\{\frac{|v_q| t}{2L} \right\}\right],
\label{sqprev}}
where the curly brackets denote the fractional part. The formula \eqref{sqprev}
is plotted with red solid lines in Fig.~\ref{fig:entmurev}, showing a good
agreement with the data, although the discrepancy increases for larger times
due to subleading contributions.

\begin{figure}[t]
\center
\begin{subfigure}{0.29\textwidth}
\includegraphics[width=\textwidth]{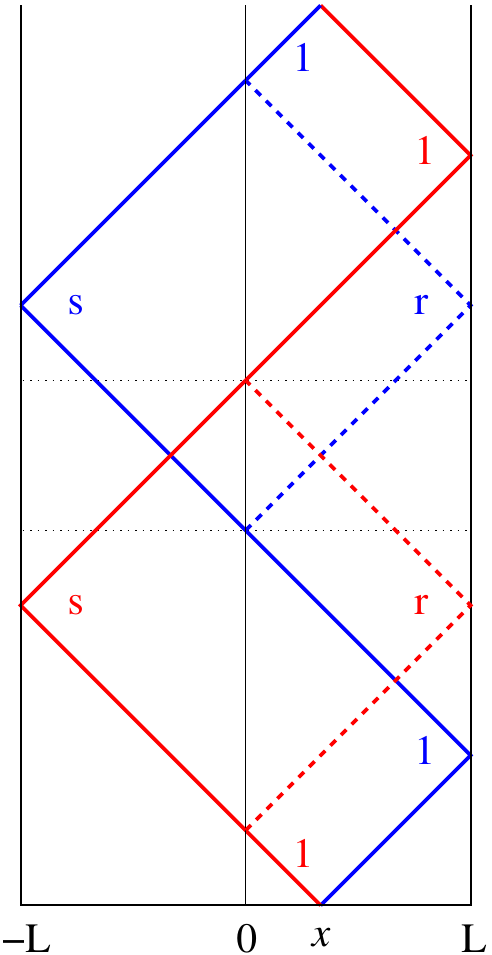}
\end{subfigure}
\begin{subfigure}{0.6\textwidth}
\qquad \includegraphics[width=.9\textwidth]{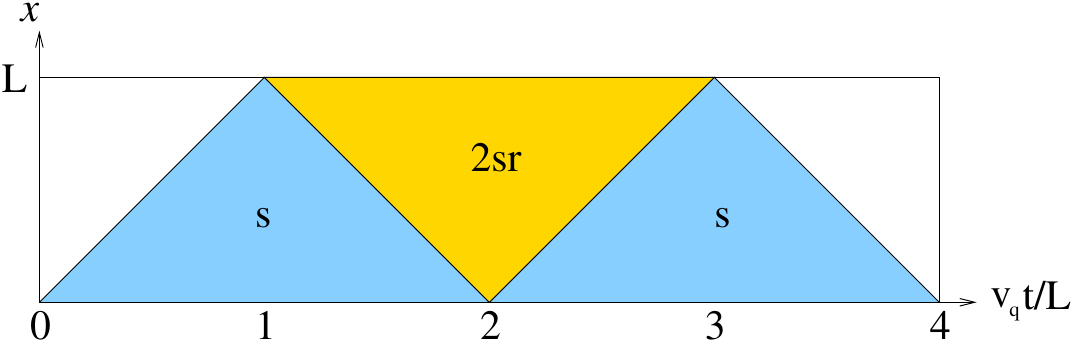}
\vskip1em
\includegraphics[width=.9\textwidth]{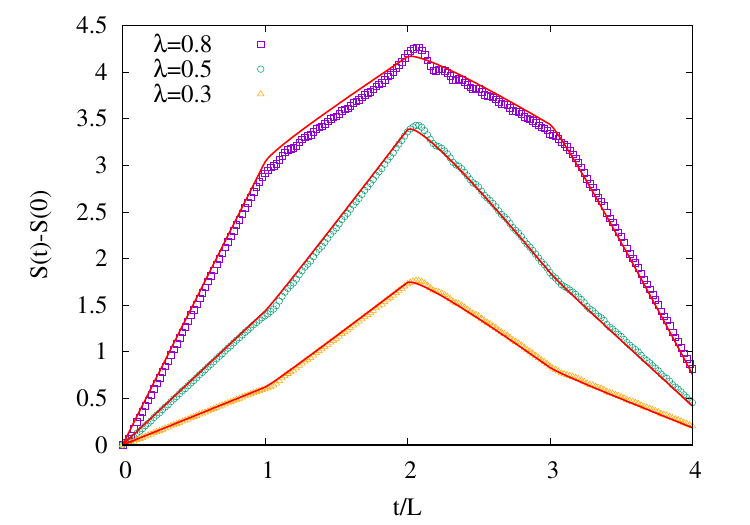}
\hskip3em
\end{subfigure}
\caption{Left: quasiparticle trajectories with corresponding amplitudes.
Top right: initial locations of quasiparticle pairs that contribute in the first full period
$0<v_qt<4L$. The shaded regions show the contributions with the amplitudes indicated.
Bottom right: entropy growth compared to the quasiparticle ansatz \eqref{sqpdefrev}.}
\label{fig:entrev}
\end{figure}

The case of the defect requires a more careful analysis, by following the quasiparticle
trajectories and scattering events for large times. This is shown on the left panel of Fig.~\ref{fig:entrev},
for a pair emitted at a distance $x$ from the defect. When reaching the defect, the transmitted
part (solid line) of the red particle picks up an amplitude $s(q)$, whereas the reflected component
(dashed line) receives an amplitude $r(q)$. For short times, the dominant contribution to the entanglement is
created by the transmitted red and the blue particles.
As discussed in sec. \ref{sec:quenchdef},
there are some correlations between the transmitted and reflected components,
however, their contribution to the entanglement is subleading for short times. 
Pictorially speaking, the dominant part of the entanglement is carried between
the red and blue lines that reside in opposite halves of the system.
For large enough times, however, the blue particle reaches the defect after a reflection
and its transmitted part (solid line) can create entanglement with the reflected part (dashed line)
of the red particle, and vice versa. These processes contribute in a time window depicted
by the horizontal dotted lines, and both of them carry an amplitude $s(q)r(q)$,
which we assume to add phase coherently. Finally, when the transmitted and reflected
beams join again, they reproduce the original wave with unit amplitude, which is due to the
fact that the amplitudes \eqref{sr} do not carry any phase and satisfy $s^2(q)+r^2(q)=1$.

The ansatz for the entropy can be found by simply summing up the contributions from
the two different processes discussed above. The range of emission points $x$ of pairs
that can contribute at a fixed time $t$ is illustrated on the top right panel of Fig.~\ref{fig:entrev}
for the first full period $0<v_qt<4L$ of time evolution. This has a triangular pattern, with the
blue regions corresponding to the process that was already discussed in sec. \ref{sec:quenchdef},
and described by the relation \eqref{nqrel} for the occupation functions. Instead, in the yellow region
one has an overall factor $2s(q)r(q)$ multiplying the correlations between the two halves of the chain,
which suggests the analogous relation
\eq{
\tilde n'_q(1- \tilde n'_q) = 4\lambda^2(1-\lambda^2) \, n_q(1-n_q) \, .
\label{nqrel2}}
In other words, the entanglement carried between transmitted and reflected wave components
is described via the modified occupation function $\tilde n'_q$. In turn, the quasiparticle ansatz reads
\begin{align}
S_n&= \int_{-\pi}^{\pi} \frac{\dd q}{2\pi} \, s_n(n'_q) \, 2L
\min \left[ \left\{\frac{|v_q| t}{2L}\right\}, 1-\left\{\frac{|v_q| t}{2L} \right\}\right] \nonumber \\
&+ \int_{-\pi}^{\pi} \frac{\dd q}{2\pi} \, s_n(\tilde n'_q) \, L
\max \left[ 0, 1-\left| 4 \left\{\frac{|v_q| t}{4L}\right\} -2 \right| \right],
\label{sqpdefrev}
\end{align}
where the first and second terms follow from carrying out the integration over $x$
in the blue and yellow regions, respectively, taking into account the periodicity of the pattern.

The ansatz \eqref{sqpdefrev} is compared against the lattice data in the bottom right panel
of Fig.~\ref{fig:entrev}, finding a remarkably good agreement. One should stress that
the pattern of entanglement is completely different from the one found in Fig.~\ref{fig:entmurev}.
Indeed, instead of the decay in the time window $L<t<2L$, one finds a continued growth of
entanglement with a slope that can even exceed the one in the initial growth phase $0<t<L$.
The decay of the entropy ensues only after $t>2L$, with a quasi-periodic pattern repeating
itself after a full period $t=4L$. The change in the behaviour is clearly due to the second type
of contributions in \eqref{sqpdefrev}. Note that for this result it is essential to consider the
initial Hamiltonian \eqref{H0} with a staggering that does not respect the reflection symmetry.
Indeed, such an initial state yields a constructive interference for the correlations carried by
the transmitted and reflected wave components, for pairs emitted on opposite sides of the defect.
In sharp contrast, for reflection symmetric staggering one obaints a destructive interference and,
as shown in appendix \ref{app:corr}, the relation $C'_{AB}=\lambda C_{AB}$ holds for
arbitrary times. In this case the second term in \eqref{sqpdefrev} is absent, and one obtains
a completely similar pattern as in Fig.~\ref{fig:entmurev}, which we verified numerically.

\subsection{Hopping defect}

\begin{figure}[htb]
\center
\includegraphics[width=0.49\textwidth]{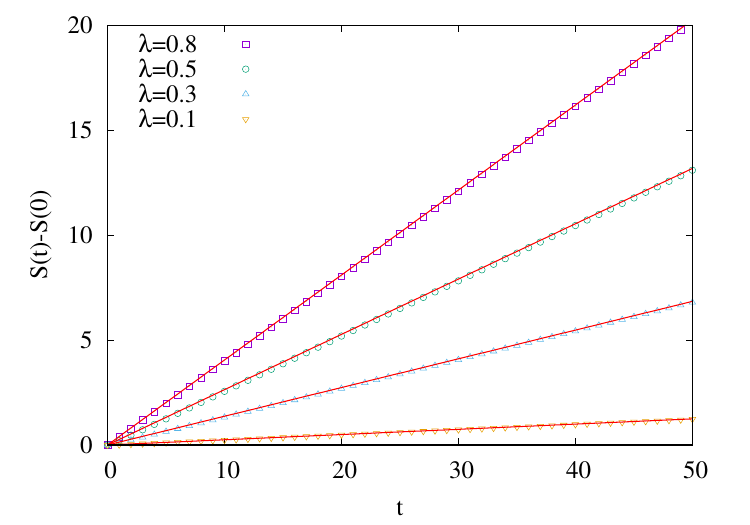}
\includegraphics[width=0.49\textwidth]{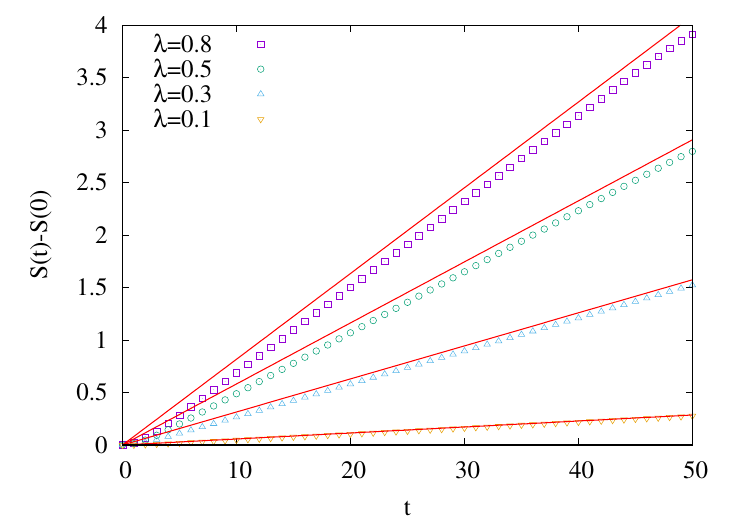}
\caption{Entropy growth for the hopping defect after a global quench with
$\mu\to\infty$ (left) and $\mu=0.1$ (right), for various $\lambda$ and $L=50$.
The red solid lines show the quasiparticle ansatz with modified
occupation function \eqref{nqphop} given via the transmission coefficient \eqref{T}.}
\label{fig:Shdef}
\end{figure}

To conclude this section, we remark that the quench results can also be generalized
to other type of defects. We restrict our attention to short times $t<L$ and, for simplicity,
we discuss the case of a hopping defect. This is given by a single weakened coupling
$H'_{0,1}=H'_{1,0}=-\lambda/2$ in the center of the chain, without the local potentials.
Setting $\lambda=\ee^{-\nu}$, the hopping defect is characterized by a transmission coefficient
\eq{
T(q) = \frac{\sin^2 q}{\sh^2 \, \nu + \sin^2 q} \, ,
\label{T}}
that depends on the momentum of the incoming particle, i.e. the defect is not any more conformal.
In the quasiparticle interpretation, this suggests that the relation \eqref{nqrel} between the occupation
functions should be modified by replacing $\lambda^2 \to T(q)$, which yields
\eq{
n'_q= \frac{1}{2} \left( 1 \pm \sqrt{1-T(q)\frac{\mu^2}{\cos^2 q+\mu^2}} \right).
\label{nqphop}}
Hence the slope of the quasiparticle ansatz \eqref{gammulam} should be evaluated with the above occupation function.
The result is compared against the numerics in Fig.~\ref{fig:Shdef} for $n=1$, with a good agreement.
Note that the main subleading term seems to be given by a constant for large times, without additional
logarithmic contributions observed for the conformal defect.

\section{Discussion}\label{sec:Conclusions}

We studied the growth of entanglement after a global quench in the presence of a conformal defect, both on the level of CFT and for a free-fermion chain. In both cases, one finds a linear growth with a slope that is modified with respect to the homogeneous case. In the CFT context, introducing the concept of boundary twist fields and scaling dimensions, we could identify the slope ratio with the effective central charge that governs the equilibrium scaling of the entropy, reproducing the results of \cite{wwr-17}. On the lattice, however, we observed slight deviations of the slope ratios, found by a proper generalization of the quasiparticle picture. The origin of the discrepancy is that our choice of initial state, prepared by adding a staggered potential, does not perfectly reproduce a thermal distribution which is assumed in the CFT treatment. Indeed, the discrepancy vanishes by considering a thermal initial state in the low-temperature limit.

For short times, the decreased slope of the half-chain entropy can be attributed to the imperfect transmission across the defect. We also studied the large time behaviour on the lattice, observing an unusual pattern of entanglement revivals. These could be understood by following the quasiparticle trajectories after bouncing back from the end of the chain and adding the entanglement contribution carried between partially reflected and transmitted components. Altogether this creates a quasi-periodic pattern of the entropy. It should be stressed, however, that an important assumption is the phase-coherent superposition of the various contributions, which requires momentum-independent transmission and reflection amplitudes. Indeed, for the hopping defect this is not satisfied and one observes a highly irregular pattern of the entropy, which is probably due to interference effects and requires further investigation.

A possible future direction is the systematic investigation of the boundary twist fields introduced in this work. In particular, the determination of the boundary scaling dimension for generic CFTs is a nontrivial task. Furthermore, a complete characterization of the correlation functions of the boundary/bulk twist fields in a boundary geometry is fundamental to access the entanglement properties for more general subsystems. We hope to come back to this problem in the future, both for a better understanding of the entanglement in the ground-state \cite{rs-21,rs-22,sths-22,mt-21,kmm-21,Levine-04,cmp-22,cmp-22a} and in other non-equilibrium protocols, as the local quench \cite{ep-12,wwr-17,Bertini-16}. In fact, in the latter case the entropy growth is logarithmic, and the prefactor is exactly given by $\ceff$ even on the lattice \cite{ep-12}. This is due to the fact, that in this low-energy quench protocol the precise form of the dispersion does not play a role.

Another interesting issue that we have left open is the question of logarithmic corrections beyond the quasiparticle ansatz. Similar corrections have been identified in the context of defect problems in \cite{fg-21,fg-22} for translational invariant steady states, and emerge due to a jump singularity in the occupation function. In our case, however, the logarithmic terms appear in the time domain and one has to deal with a non-translational invariant setting. Furthermore, some logarithmic corrections seem to appear even in the homogeneous quench setting (see Fig.~\ref{fig:entmu}), which must have some different origin and is left for future studies.

A further natural extension of our work is to study bosonic chains with defects. While the CFT derivation still applies, with an effective central charge obtained in \cite{ss-08}, the quasiparticle ansatz is expected to be different from the fermionic case. One could also investigate other interacting models, such as the XXZ chain, where entanglement properties can be significantly influenced by the presence of defects \cite{zpw-06,cc-13,bm-17,kmv-14}. It would be interesting to see whether the quasiparticle picture can be extended also for those systems.

\section*{Acknowledgements}

We would like to thank Pasquale Calabrese for discussions.
VE acknowledges funding from the Austrian Science Fund (FWF) through Project No. P35434-N. LC acknowledges support from ERC under Consolidator grant number 771536 (NEMO).

\begin{appendix}

\section{Diagonalization of the staggered chain\label{app:stag}}

Here we show how to diagonalize the pre-quench Hamiltonian \eqref{H0} for homogeneous hoppings ($\lambda=1$),
and calculate the occupation function of the post-quench Hamiltonian in the ground state of $\hat H_0$.
We are interested in the thermodynamic limit, and thus assume periodic boundary conditions for simplicity.
The Hamiltonian is then of size $2L$ and two-site translation invariant.
Let us introduce the operators $a_m=c_{2m-1}$ and  $b_m=c_{2m}$ and rewrite the Hamiltonian as
\eq{
\hat H_0 = -\frac{1}{2}\sum_{m=1}^{L} (a^\dag_m b_m + b^\dag_m a_{m+1} + \textrm{h.c.})
- \mu \sum_m (a^\dag_m a_m - b^\dag_m b_m)
}
We first perform a Fourier transformation on both sublattices
\eq{
\left(\begin{array}{c} a_m \\ b_m \end{array}\right) =
\frac{1}{\sqrt{L}} \sum_p \ee^{ipm}
\left(\begin{array}{c} a_p \\ b_p \end{array}\right) ,
\qquad p=\frac{2\pi}{L}k \, ,
\label{FTsub}}
where $p$ is the sublattice momentum and its index runs over $k = -L/2+1, \dots, L/2$.
Introducing new operators via
\begin{align}
&a_p=\frac{1}{\sqrt{2}}
\left( \sqrt{1+\frac{\mu}{\Omega_p}} \alpha_p 
+ \sqrt{1-\frac{\mu}{\Omega_p}} \ee^{-i\frac{p}{2}} \beta_p \right)
\\
&b_p=\frac{1}{\sqrt{2}}
\left( \sqrt{1-\frac{\mu}{\Omega_p}} \ee^{i\frac{p}{2}} \alpha_p 
- \sqrt{1+\frac{\mu}{\Omega_p}}  \beta_p \right)
\label{aqbq}
\end{align}
one arrives at the diagonal form of the Hamiltonian
\eq{
H = -\sum_p \Omega_p (\alpha_p^\dag \alpha_p - \beta_p^\dag \beta_p) \, ,
\qquad
\Omega_p = \sqrt{\cos^2 \frac{p}{2}+\mu^2} \, .
\label{hdiag}}
The half-filled ground state is given by all the $\alpha_p$ particles occupied
$\braket{\alpha^\dag_p\alpha_p}=1$, whereas the $\beta_p$ band is empty,
$\braket{\beta^\dag_p\beta_p}=0$.

The occupation numbers $n_q = \braket{c^\dag_q c_q}$ for the homogeneous
post-quench Hamiltonian are defined in terms of the Fourier modes on the full chain
\eq{
c_q  = \frac{1}{\sqrt{2L}} \sum_{j=1}^{2L} \ee^{-iqj} c_j \, ,
\quad q=\frac{2\pi}{2L}k \, ,
}
where the index runs over $k = -L+1, \dots, L$. One has then
\eq{
n_q = \frac{1}{2L} \sum_{m,n=1}^{L} \left[ \ee^{iq(2m-2n)}
(\braket{a^\dag_m a_n} + \braket{b^\dag_m b_n} ) +
\ee^{iq(2m-2n-1)} \braket{a^\dag_m b_n}+
\ee^{iq(2m-2n+1)} \braket{b^\dag_m a_n} \right].
\label{nq1}}
The ground-state expectation values can be evaluated using \eqref{FTsub} and \eqref{aqbq} as
\eq{
\braket{a^\dag_m a_n} = \frac{1}{L}\sum_p \ee^{-ip(m-n)} \frac{1}{2}\left(1+\frac{\mu}{\Omega_p}\right), \qquad
\braket{b^\dag_m b_n} = \frac{1}{L}\sum_p \ee^{-ip(m-n)} \frac{1}{2}\left(1-\frac{\mu}{\Omega_p}\right),
\label{aabb}}
as well as
\eq{
\braket{a^\dag_m b_n} = \frac{1}{L}\sum_p \ee^{-ip(m-n-1/2)} \frac{1}{2} \sqrt{1-\frac{\mu^2}{\Omega^2_p}} \, , \qquad
\braket{b^\dag_m a_n} = \frac{1}{L}\sum_p \ee^{-ip(m-n+1/2)} \frac{1}{2}\sqrt{1-\frac{\mu^2}{\Omega^2_p}} \, .
\label{abba}}
Let us first consider the $q$ modes restricted to the reduced Brillouin zone $k = -L/2+1, \dots, L/2$
of the sublattice. Substituting \eqref{aabb} and \eqref{abba} into \eqref{nq1}, the sums over $m$ and $n$
both give a factor $L \, \delta_{p,2q}$ and one arrives at
\eq{
n_q = \frac{1}{2}\left(1+\frac{|\cos q|}{\sqrt{\cos^2 q+\mu^2}}\right).
\label{nq2}}
On the other hand, for $q$ values with $k=-L+1,\dots,-L/2$ and $k=L/2+1,\dots,L$ outside the
reduced Brillouin zone, the $k$ index must be shifted to match the $p$ momentum,
and the sums over $m$ and $n$ yield $L \, \delta_{p,2q\pm 2\pi}$, respectively.
The extra phase factors $\ee^{\pm i(p/2-q)}$ appearing in the second and third term of
\eqref{nq1} thus yield an extra sign and one has
\eq{
n_q = \frac{1}{2}\left(1-\frac{|\cos q|}{\sqrt{\cos^2 q+\mu^2}}\right).
\label{nq3}}
However, in this regime the $\cos q$ function is negative and thus the two
solutions \eqref{nq2} and \eqref{nq3} give an analytic function in the full Brillouin zone
as reported in \eqref{nqmu} in the main text.

\section{Calculation of the slope ratio for $\mu \to 0$\label{app:mu0}}

In this appendix we analyze the slope of the R\'enyi entropy (see Eq. \eqref{gammulam})
\be
\gamma_n(\mu,\lambda) = \int^{\pi}_{-\pi} \frac{dq}{2\pi} |\sin q| s_n\left( \frac{1}{2}\left( 1-\sqrt{1-\lambda^2 \frac{\mu^2}{\cos^2 q+\mu^2}} \right) \right)
\label{eq:slope_mu0}
\eeq
in the limit $\mu \rightarrow 0^+$, which requires a careful analysis. A preliminary observation is that for $\mu=0$ one gets
\be
\gamma_n(0,\lambda) = \int^{\pi}_{-\pi} \frac{dq}{2\pi} |\sin q| s_n(0)=0,
\eeq
which is a correct conclusion, since in this limit the pre/post-quench Hamiltonians are the same and there is no dynamics. However, here we are mostly interested in the way $\gamma_n(\mu,\lambda)$ goes to zero as $\mu\rightarrow 0$, to extract eventually the finite limit
\be
\underset{\mu \rightarrow 0}{\lim} \frac{\gamma_n(\mu,\lambda)}{\gamma_n(\mu,1)},
\eeq
which we denote by $\frac{\gamma_n(0,\lambda)}{\gamma_n(0,1)}$ with a slight abuse of notation.

To proceed with the evaluation of \eqref{eq:slope_mu0}, we first notice that the nontrivial contributions to the integral come from the values of the momentum $q$ close to $\pm \pi/2$, an observation which motivates the change of variable $q\rightarrow q+\pi/2$, and we get
\be
\gamma_n(\mu,\lambda) = \int^{\pi/2}_{-\pi/2} \frac{dq}{\pi} \cos q \ s_n\left( \frac{1}{2}\left( 1-\sqrt{1-\lambda^2 \frac{\mu^2}{\sin^2 q+\mu^2}} \right) \right).
\eeq
At this point, it is natural to introduce the scaling variable
\be
z \equiv \frac{\sin q}{\mu},
\eeq
so that we can write the slope as
\be
\label{eq:slope_mu0_bis}
\gamma_n(\mu,\lambda) = \mu \int^{1/\mu}_{-1/\mu} \frac{dz}{\pi} \ s_n\left( \frac{1}{2}\left( 1-\sqrt{1-\lambda^2 \frac{1}{1+z^2}} \right) \right).
\eeq
Up to now, no approximation has been done and Eq. \eqref{eq:slope_mu0_bis} is exact for any finite value of $\mu$. Nevertheless, the advantage of this expression comes from the fact that the singular behaviour of the integrand in Eq. \eqref{eq:slope_mu0} is not present anymore in Eq. \eqref{eq:slope_mu0_bis} , which makes the last expression suitable for a numerical evaluation.

From now on, we consider explicitly the limit of small $\mu$ and we extract only the term of order $O(\mu)$ in Eq. \eqref{eq:slope_mu0_bis}, obtaining
\be
\frac{\gamma_n(\mu,\lambda)}{\mu} \simeq \int^{\infty}_{-\infty} \frac{dz}{\pi} \ s_n\left( \frac{1}{2}\left( 1-\sqrt{1-\lambda^2 \frac{1}{1+z^2}} \right) \right),
\label{eq:slope_mu0_tris}
\eeq
which is the main result of this appendix. While we were not able to perform analytically the integral \eqref{eq:slope_mu0_tris} for any real value of $n$, and in particular for $n=1$ directly related to the entanglement entropy, we provide below a simple closed expression for integer $n\geq 2$. The key observation is the following decomposition of the density of R\'enyi entropy
\be
s_n(x) \equiv \frac{1}{1-n}\log(x^n+(1-x)^n) = \frac{1}{1-n}\sum^{(n-1)/2}_{p=-(n-1)/2} \log( e^{i2\pi p /n}x+(1-x)),
\label{eq:sn_decomp}
\eeq
a simple identity which comes from the factorization of the polynomial $x^n+(1-x)^n$. We introduce for convenience the function $f(\alpha)$ as
\be
\begin{split}
f(\alpha) \equiv \frac{1}{2} \int^{\infty}_{-\infty} \frac{dz}{\pi} \log \left(
\frac{1-\sqrt{1-\lambda^2\frac{1}{1+z^2}}}{2} e^{i\alpha} + \frac{1+\sqrt{1-\lambda^2\frac{1}{1+z^2}}}{2}    \right)+\\
\frac{1}{2} \int^{\infty}_{-\infty} \frac{dz}{\pi} \log \left(
\frac{1-\sqrt{1-\lambda^2\frac{1}{1+z^2}}}{2} e^{-i\alpha} + \frac{1+\sqrt{1-\lambda^2\frac{1}{1+z^2}}}{2}    \right),
\end{split}
\eeq
related to $\gamma_n(\mu,\lambda)$ by the following relation
\be
\frac{\gamma_n(\mu,\lambda)}{\mu} = \frac{1}{1-n}\sum^{(n-1)/2}_{p=-(n-1)/2} f(2\pi p/n),
\label{eq:gamma_falpha}
\eeq
which is a straightforward consequence of Eq. \eqref{eq:sn_decomp}. Interestingly, one is able to compute $f(\alpha)$ analytically, as we will show, and then a closed expression for the slope $\gamma_n(\mu,\lambda)$ for integer $n\geq 2$ can be provided. To do so, we first differentiate $f(\alpha)$ over $\alpha$ and we get
\be
\frac{d}{d\alpha} f(\alpha) = -\int^{\infty}_{-\infty} \frac{dz}{\pi}\frac{\lambda^2 \sin (\alpha )}{2 \lambda^2 \cos (\alpha )+4 z^2-2\lambda^2+4} = -\frac{\lambda^2 \sin (\alpha )}{2 \sqrt{2 \lambda^2 \cos (\alpha )-2 \lambda^2+4}}.
\eeq
Then, we reintegrate back and obtain
\be
f(\alpha) = \sqrt{1-\lambda^2\sin^2 ( \alpha/2)}-1,
\label{eq:falpha_value}
\eeq
where the integration constant has been chosen so that $f(0)=0$, a property which follows from the definition of $f(\alpha)$. Inserting \eqref{eq:falpha_value} into \eqref{eq:gamma_falpha} we finally arrive at the expression \eqref{gammu0} reported in the main text.

We conclude this appendix with the computation of the slopes $\bar{\gamma}_n(\beta,\lambda)$ for an initial thermal state, highlighting its discrepancies with  $\gamma_n(\mu,\lambda)$. The occupation number of such state at temperature $\beta^{-1}$ is
\be
\bar{n}_q = \frac{1}{e^{-\beta \cos q}+1},
\eeq
and the quasiparticle ansatz of the slope gives (see Eq. \eqref{gammulam})
\be
\bar{\gamma}_n(\beta,\lambda) = \int^{\pi}_{-\pi}\frac{dq}{2\pi} |\sin q|s_n\left( \frac{1-\sqrt{1-4\lambda^2\bar{n}_q(1-\bar{n}_q)}}{2} \right).
\eeq
While the latter expression is valid for any $\beta$, we are mostly interested in the low-temperature limit ($\beta \rightarrow \infty$), whose features are expected to be captured by CFT. After simple algebra, coming from the linearization of the dispersion around the Fermi-point $q=\pi/2$, one gets
\be\label{eq:slope_CFT}
\bar{\gamma}_n(\beta,\lambda) \simeq 2\int^{\infty}_{0}\frac{dq}{\pi} s_n\left( \frac{1-\sqrt{1-\frac{\lambda^2}{\cosh^2 \frac{\beta q}{2}}}}{2} \right) = \frac{2}{\beta} \int^{\infty}_{0}\frac{dx}{\pi} s_n\left( \frac{1-\sqrt{1-\frac{\lambda^2}{\cosh^2 \frac{x}{2}}}}{2} \right),
\eeq
a relation valid in the limit of large $\beta$. A tedious but straightforward calculation of the integral in Eq. \eqref{eq:slope_CFT}, analogous to the computation of $\gamma_n(\mu,\lambda)$, gives for integer $n\geq 2$
\be
\bar{\gamma}_n(\beta,\lambda) = \frac{1}{\beta}\frac{2\pi}{n-1} \sum^{(n-1)/2}_{p=-(n-1)/2} \left(\frac{1}{\pi}\arcsin\left(\lambda \sin \frac{\pi p}{n}\right)\right)^2.
\eeq
If one compares the latter formula with the logarithmic prefactor in Eq. \eqref{kappan}, valid for the ground-state entanglement (both in CFT and on the lattice), one realizes that they are proportional. This can be regarded as a non-trivial consistency check of the quasiparticle ansatz with the CFT results in the presence of a defect.

\section{Correlation matrix for reflection-symmetric staggered potential\label{app:corr}}

In this appendix we consider a quench from the ground state of a chain with a conformal defect
and an additional staggered field that is symmetric under reflection
\eq{
\hat H_0 = \hat H + \mu \sum_{n=1}^{L} (-1)^n c^\dag_n c_n-\mu \sum_{n=-L+1}^{0} (-1)^n c^\dag_n c_n \, .
}
Note that the defect Hamiltonian $\hat H$ in \eqref{Hdef} and thus the full Hamiltonian does not have a reflection symmetry. However, in the defect-free ($\lambda=1$) case one does have a reflection symmetry, which is enough to show that the single-particle eigenvectors and eigenvalues of $\hat H_0$ possess the very same property \eqref{phikdef} as those without the staggered field. Namely, one has
\eq{
\tilde\phi'_k(m) = \left\{
\begin{array}{ll}
\alpha_k \tilde \phi_k(m) & \quad  m \le 0 \\
\beta_k \tilde \phi_k(m) & \quad  m \ge 1
\end{array}\right. , \qquad
\tilde \Omega'_k = \tilde \Omega_k \, ,
\label{tphikdef}
}
with the coefficients given in \eqref{ab}.
Note that we do not have a closed form expression for $\tilde \phi_k(m)$,
but we will not need it. The only property we need is that, due to the reflection
symmetry of $\hat H_0$ at $\lambda=1$, the eigenvectors satisfy $\tilde \phi_k(1-m)=(-1)^{k-1} \tilde \phi_k(m)$.

We now calculate the time-dependent correlation matrix after the quench
\eq{
C'(t) = \ee^{iH't}C'(0)\ee^{-iH't} \, ,
}
where the post-quench Hamiltonian is the same as in the main text. Defining
\eq{
A'_{k,l}= \sum_{i,j} \phi'_k(i) \, C'_{i,j}(0) \, \phi'_l(j) \, ,
}
one has for the matrix elements
\eq{
C'_{m,n} = \sum_{k,l}  \phi'_k(m) \, A'_{k,l} \, \phi'_l(n) \, \ee^{i(\omega'_k-\omega'_l)t} \, .
}
We are interested in the off-diagonal part ($m\le 0$ and $n\ge 1$), where from \eqref{phikdef} one obtains
\eq{
C'_{m,n} = \sum_{k,l} \alpha_k \beta_l \, \phi_k(m) \, A'_{k,l} \, \phi_l(n) \, \ee^{i(\omega_k-\omega_l)t} \, .
\label{Cpmn}}
Furthermore, one can rewrite $A'_{k,l}$ in terms of an overlap matrix
\eq{
A'_{k,l}= \sum_{p \in F} B'_{k,p}B'_{l,p} \, , \qquad
B'_{k,p} = \sum_{i=-L+1}^{L} \phi'_k(i) \tilde\phi'_p(i) \, ,
}
where the sum goes over the Fermi sea $F$, satisfying $\tilde\Omega_p<0$.
Using \eqref{tphikdef}, the overlap matrix can be written as
\eq{
B'_{k,p} = 
\alpha_k \alpha_p \sum_{i \le 0} \phi_k(i) \tilde\phi_p(i)+
\beta_k \beta_p \sum_{i \ge 1} \phi_k(i) \tilde\phi_p(i) \, .
}

To proceed, let us first note the following identitites
\eq{
\alpha_k \alpha_l=
\begin{cases}
\alpha_k^2 & \textrm{for $k-l$ even} \\
\lambda & \textrm{for $k-l$ odd}
\end{cases}
\, , \qquad 
\beta_k \beta_l = 
\begin{cases}
\beta_k^2  & \textrm{for $k-l$ even} \\
\lambda & \textrm{for $k-l$ odd}
\end{cases} \, .
\label{akal}}
Using the symmetry properties of both eigenvectors $\phi_k$ and $\tilde \phi_p$, one arrives at
\eq{
B'_{k,p} = 
(\beta_k \beta_p + (-1)^{k+p} \alpha_k \alpha_p) \sum_{i \ge 1} \phi_k(i) \tilde\phi_p(i)=
B_{k,p} \, ,
}
which follows via \eqref{akal}, from the fact that the prefactor of the sum is nonvanishing
only for even $k-p$, where it evaluates to $\alpha^2_k+\beta^2_k=2$.
The overlap matrix is thus completely independent of $\lambda$ and its matrix elements are
nonvanishing only for $k-p$ even. This implies that $A'_{k,l}=A_{k,l}$
with nonvanishing elements for $k-l$ even. Finally, for $k-l$ even one has $\alpha_k\beta_l=\lambda$,
and thus \eqref{Cpmn} yields the required identity $C'_{m,n}=\lambda C_{m,n}$ for the off-diagonal
elements.

\end{appendix}

\bibliography{bibliography.bib}

\nolinenumbers

\end{document}